\RequirePackage{etoolbox}
\csdef{input@path}{%
 {sty/}
 {img/}
}%

\documentclass[generic]{imsart}
\usepackage{amsthm}
\usepackage{amsmath}
\usepackage{natbib}
\usepackage[colorlinks,citecolor=blue,urlcolor=blue,filecolor=blue,backref=page]{hyperref}
\usepackage{graphicx}
\usepackage{cancel}
\usepackage{subfigure}
\usepackage{epstopdf}
\usepackage{geometry}
\usepackage{graphics}
\usepackage{booktabs}

\startlocaldefs
\newcommand{\given}{\,|\,}
\newcommand{\T}{\top}

\newcommand{\calS}{{\cal S}}
\newcommand{\calR}{{\cal R}}
\newcommand{\calL}{{\cal L}}

\newcommand{\calT}{{\cal T}}
\newcommand{\calU}{{\cal U}}
\newcommand{\calV}{{\cal V}}

\newcommand{\taus}{\tau^2}

\newcommand{\tildew}{\tilde{w}}
\newcommand{\tildeK}{\tilde{K}}

\newcommand{\ra}[1]{\renewcommand{\arraystretch}{#1}}

\endlocaldefs

\begin{document}


\begin{frontmatter}
\title{High-dimensional Bayesian Geostatistics}

\runtitle{High-dimensional Bayesian Geostatistics}

\begin{aug}
\author{\fnms{Sudipto} \snm{Banerjee}\ead[label=e1]{sudipto@ucla.edu}}

\runauthor{Sudipto Banerjee}

\address[]{UCLA Department of Biostatistics\\ 650 Charles E. Young Drive South\\ Los Angeles, CA 90095-1772.}


\end{aug}

\begin{abstract}
With the growing capabilities of Geographic Information Systems (GIS) and user-friendly software, statisticians today routinely encounter geographically referenced data containing observations from a large number of spatial locations and time points. Over the last decade, hierarchical spatiotemporal process models have become widely deployed statistical tools for researchers to better understand the complex nature of spatial and temporal variability. However, fitting hierarchical spatiotemporal models often involves expensive matrix computations with complexity increasing in cubic order for the number of spatial locations and temporal points. This renders such models unfeasible for large data sets. This article offers a focused review of two methods for constructing well-defined highly scalable spatiotemporal stochastic processes. Both these processes can be used as ``priors" for spatiotemporal random fields. The first approach constructs a low-rank process operating on a lower-dimensional subspace. The second approach constructs a Nearest-Neighbor Gaussian Process (NNGP) that ensures sparse precision matrices for its finite realizations. Both processes can be exploited as a scalable prior embedded within a rich hierarchical modeling framework to deliver full Bayesian inference. These approaches can be described as model-based solutions for big spatiotemporal datasets. The models ensure that the algorithmic complexity has $\sim n$ floating point operations (flops), where $n$ the number of spatial locations (per iteration). We compare these methods and provide some insight into their methodological underpinnings.
\end{abstract}


\end{frontmatter}


\section{Introduction}\label{sec:Intro}
\noindent The increased availability of inexpensive, high speed computing has enabled the collection of massive amounts of spatial and spatiotemporal datasets across many fields. This has resulted in widespread deployment of sophisticated Geographic Information Systems (GIS) and related software, and the ability to investigate challenging inferential questions related to geographically-referenced data. 
See, for example, the books by \cite{cres93}, \cite{stein99}, \cite{moll03}, \cite{scha04}, \cite{geldigfuegut}, \cite{creswikle11} and \cite{ban14} for a variety of statistical methods and applications.  

This article will focus only on point-referenced data, which refers to data referenced by points with coordinates (latitude-longitude, Easting-Northing etc.). Modeling typically proceeds from a spatial or spatiotemporal process that introduces dependence among any finite collection of random variables from an underlying random field. For our purposes, we will consider the stochastic process as an uncountable set of random variables, say $\{w(\ell): \ell\in \calL\}$, over a domain of interest $\calL$, which is endowed with a probability law specifying the joint distribution for any finite sample from that set. For example, in spatial modeling $\calL$ is often assumed to be a subset of points in the Euclidean space $\Re^d$ (usually $d=2$ or $3$) or, perhaps, a set of geographic coordinates over a sphere or ellipsoid. In spatiotemporal settings $\calL = \calS\times \calT$, where $\calS \subset \Re^d$ is the spatial region, $\calT \subset [0,\infty)$ is the time domain and $\ell = (s,t)$ is a space-time coordinate with spatial location $s \in \calS$ and time point $t\in \calT$ \citep[see, e.g.,][for details]{gnei10}. 

Such processes are specified with a \emph{covariance function} $K_{\theta}(\ell, \ell')$ that gives the covariance between $w(\ell)$ and $w(\ell')$ for any two points $\ell$ and $\ell'$ in $\calL$. For any finite collection $\calU=\{ \ell_1, \ell_2, \ldots, \ell_n \}$ in $\calL$, let $w_\calU = (w(\ell_1)), w(\ell_2), \ldots, w(\ell_n))^{\T}$ be the realizations of the process over $\calU$. Also, for two finite sets $\calU$ and $\calV$ containing $n$ and $m$ points in $\calL$, respectively, we define the $n\times m$ matrix $K_{\theta}(\calU,\calV)=\mbox{Cov}(w_\calU,w_\calV \given \theta)$, where the covariances are evaluated using $K_{\theta}(\cdot, \cdot)$. When $\calU$ or $\calV$ contains a single point, $K_{\theta}(\calU,\calV)$ is a row or column vector, respectively. A valid spatiotemporal covariance function ensures that $K_{\theta}(\calU,\calU)$ is positive definite for any finite set $\calU$. In geostatistics, we usually deal with a fixed set of points $\calU$ and, if the context is clear, we write $K_{\theta}(\calU,\calU)$ simply as $K_{\theta}$. A popular specification assumes $\{w(\ell): \ell\in \calL\}$ is a zero-centered Gaussian process written as $w(\ell)\sim GP(0, K_{\theta}(\cdot,\cdot))$, which implies that the $n\times 1$ vector $w = (w(\ell_1), w(\ell_2)\ldots, w(\ell_n))^{\T}$ is distributed as $N(0, K_{\theta})$, where $K_{\theta}$ is the $n\times n$ covariance matrix with $(i,j)$-th element $K_{\theta}(\ell_i,\ell_j)$. 
Various characterizations and classes of valid spatial (and spatiotemporal) covariance functions can be found in \cite{gnei10}, \cite{cres93}, \cite{stein99}, \cite{geldigfuegut}, \cite{creswikle11} and \cite{ban14} and numerous references therein. The more common assumptions are of \emph{stationarity} and \emph{isotropy}. The former assumes that $K_{\theta}(\ell,\ell')=K_{\theta}(\ell-\ell')$ depends upon the coordinates only through their separation vector, while isotropy goes a step further and assumes the covariance is a function of the distance between them.  

Spatial and spatiotemporal processes are conveniently embedded within Bayesian hierarchical models. The most common geostatistical setting assumes a response or dependent variable $y(\ell)$ observed at a generic point $\ell$ along with a $p\times 1$ ($p<n$) vector of spatially referenced predictors $x(\ell)$. Model-based geostatistical data analysis customarily envisions a spatial regression model,
\begin{equation}\label{eq: BasicModel}
y(\ell) = x^{\T}(\ell)\beta + w(\ell) + \epsilon(\ell)\; , 
\end{equation}
where $\beta$ is the $p\times 1$ vector of slopes, and the residual from the regression is the sum of a spatial or spatiotemporal process, $w(\ell)\sim GP(0, K_{\theta}(\cdot,\cdot))$ capturing spatial and/or temporal association, and an independent process, $\epsilon(\ell)$ 
modeling measurement error or fine scale variation attributed to disturbances at distances smaller than the minimum observed separations in space and time. A Bayesian spatial model can now be constructed from (\ref{eq: BasicModel}) as
\begin{align}\label{eq: Bayesian_Spatial_Gaussian_Generic}
p(\theta, \beta, \tau) \times N(w\given 0, K_{\theta}) \times N(y \given X\beta + w, D_{\tau})\; , 
\end{align}
where $y = (y(\ell_1),y(\ell_2),\ldots,y(\ell_n))^{\T}$ is the $n\times 1$ vector of observed outcomes, $X$ is the $n\times p$ matrix of regressors with $i$-th row $x^{\T}(\ell_i)$ and the noise covariance matrix $D(\tau)$ represents measurement error or micro-scale variation and depends upon a set of variance parameters $\tau$. A common specification is $D_{\tau} = \tau^2I_n$, where $\tau^2$ is called the ``nugget.'' The hierarchy is completed by assigning prior distributions to $\beta$, $\theta$ and $\tau$. 

Bayesian inference can proceed by sampling from the joint posterior density in (\ref{eq: Bayesian_Spatial_Gaussian_Generic}) using, for example, Markov chain Monte Carlo (MCMC) methods \citep[see, e.g.,][]{robecase04}. A major computational bottleneck emerges from the size of $K_{\theta}$ in computing (\ref{eq: Bayesian_Spatial_Gaussian_Generic}). Since $\theta$ is unknown, each iteration of the model fitting algorithm will involve decomposing  or factorizing $K_{\theta}$, which typically requires $\sim n^3$ floating point operations (flops). Memory requirements are of the order $\sim n^2$. These become prohibitive for large values of $n$ when $K_{\theta}$ has no exploitable structure. Evidently, multivariate process settings, where $y(\ell)$ is a $q\times 1$ vector of outcomes, exacerbate the computational burden by a factor of $q$. For Gaussian likelihoods, one can integrate out the random effects $w$ from (\ref{eq: Bayesian_Spatial_Gaussian_Generic}). This reduces the parameter space to $\{\tau^2,\theta,\beta\}$, but one still needs to work with $K_{\theta} + \tau^2I_n$, which is again $n\times n$. These settings are referred to as ``big-n'' or ``high-dimensional'' problems in geostatistics and are widely encountered in environmental sciences today.

As modern data technologies are acquiring and exploiting massive amounts of spatiotemporal data, modeling and inference for large spatiotemporal datasets are receiving increased attention. In fact, it is impossible to provide a comprehensive review of all existing methods for geostatistical models for massive spatial data sets; \cite{sunligenton11} offers an excellent review for a number of methods for high-dimensional geostatistics. The ideas at the core of fitting models for large spatial and spatiotemporal data concern effectively solving positive definite linear systems such as $Ax=b$, where $A$ is a covariance matrix. Thus one can use probability models to build computationally efficient covariance matrices. One approach is to approximate or model $A$ with a covariance structure that can significantly reduce the computational burden. An alternative is to model $A^{-1}$ itself with an exploitable structure so that the solution $A^{-1}b$ is available without computing the inverse. For full Bayesian inference, one also needs to ensure that the determinant of $A$ is available easily.       


We remark that when inferring about stochastic processes, it is also possible to work in the spectral domain. This rich, and theoretically attractive, option has been advocated by \cite{stein99} and \cite{fuentes07} and completely avoids expensive matrix computations. The underlying idea is to transform to the space of frequencies, construct a periodogram (an estimate of the spectral density), and exploit the Whittle likelihood \citep[see, e.g.,][]{whittle54,guyon95} in the  spectral domain as an approximation to the data likelihood in the original domain. The Whittle likelihood\index{Whittle likelihood} requires no matrix inversion so, as a result, computation is very rapid. In principle, inversion back to the original space is straightforward. However, there are practical impediments. First, there is discretization to implement a fast Fourier transform whose performance can be tricky over large irregular domains. Predictive inference at arbitrary locations also will not be straightforward. Other issues include arbitrariness to the development of a periodogram.  Empirical experience is employed to suggest how many low frequencies should be discarded.  Also, there is concern regarding the performance of the Whittle likelihood as an approximation to the exact likelihood. While this approximation is reasonably well centered, it does an unsatisfactory job in the tails (thus leading to poor estimation of model variances). Lastly, modeling non-Gaussian first stages will entail unobservable random spatial effects, making the implementation impossible. In summary, use of the spectral domain with regard to handling large $n$, while theoretically attractive, has limited applicability.  
 
Broadly speaking, model-based approaches for large spatial datasets proceeds from either exploiting ``low-rank” models or exploiting ``sparsity''. The former attempts to construct Gaussian processes on a lower-dimensional subspace \cite[see, e.g., ][]{wikle99,hig01,kam03,quinonerorasmussen05,stein07,gramlee2008,stein08,cres08,ban08,cra08,sanso08,fin09,lemossanso09,cres10} in spatial, spatiotemporal and more general Gaussian process regression settings. Sparse approaches include covariance tapering \citep[see, e.g.,][]{fur06,kauf08,du09,shabytaper} using compactly supported covariance functions. This is effective for parameter estimation and interpolation of the response (``kriging''), but it has not been fully evaluated for fully Bayesian inference on residual or latent processes. Introducing sparsity in $K_{\theta}^{-1}$ is prevalent in approximating Gaussian process likelihoods using Markov random fields \citep[e.g.,][]{rueheld04}, products of lower dimensional conditional distributions \citep{ve88,ve92,stein04}, or composite likelihoods \citep[e.g.,][]{bevilacqua14,eidsvik14}.  


This article aims to provide a focused review of some massively scalable Bayesian hierarchical models for spatiotemporal data. The aim is not to provide a comprehensive review of all existing methods. Instead, we focus upon two fully model-based approaches that can be easily embedded within hierarchical models and deliver full Bayesian inference. These are low-rank processes and sparsity-inducing processes. Both these processes can be used as ``priors" for spatiotemporal random fields. Here is a brief outline of the paper. Section~\ref{sec:low_rank} discusses a Bayesian hierarchical framework for low-rank models and their implementation. Section~\ref{sec:sparsity} discusses some recent developments in sparsity-inducing Gaussian processes, especially nearest-neighbor Gaussian processes, and their implementation. Finally, Section~\ref{sec:future_directions} provides a brief account of outstanding issues for future research.     

\section{Hierarchical low-rank models}\label{sec:low_rank}

\noindent A popular way of dealing with large spatial datasets is to devise models that bring about dimension reduction \citep[][]{wikle99}. A \emph{low rank} or \emph{reduced rank} specification is typically based upon a representation or approximation in terms of the realizations of some latent process over a smaller set of points, often referred to as \emph{knots}. To be precise,
\begin{equation}\label{eq:generic_low_rank}
w(\ell) \approx \tilde{w}(\ell) = \sum_{j=1}^{r} b_{\theta}(\ell,\ell_{j}^{*}) z(\ell_{j}^{*}) = b_{\theta}^{\T}(\ell)z,
\end{equation}
where $z(\ell)$ is a well-defined process and $b_{\theta}(s,s')$ is a family of basis functions possibly depending upon some parameters $\theta$. The collection of $r$ locations $\{\ell_1^*,\ell^*_2,\ldots,\ell^*_r\}$ are the knots, $b_{\theta}(\ell)$ and $z$ are $r\times 1$ vectors with components $b_{\theta}(\ell,\ell_{j}^{*})$ and $z(\ell_j^*)$, respectively. For any collection of $n$ points, the $n\times 1$ vector $\tilde{w} = (\tilde{w}(\ell_1), \tilde{w}(\ell_2),\ldots,\tilde{w}(\ell_{n}))^{\T}$ is represented as $\tildew = B_{\theta}z$, where $B_{\theta}$ is $n\times r$ with $(i,j)$-th element $b_{\theta}(\ell_i,\ell_{j}^{*})$. Irrespective of how big $n$ is, we now have to work with the $r$ (instead of $n$) $z(\ell_j^*)$'s and the  $n\times r$ matrix $B_{\theta}$. Since we anticipate $r << n$, the consequential dimension reduction is evident and, since we will write the model in terms of the $z$'s (with the $\tilde{w}$'s being deterministic from the $z$'s, given $b_{\theta}(\cdot,\cdot)$), the associated matrices we work with will be $r \times r$.  Evidently, $\tilde{w}(\ell)$ as defined in (\ref{eq:generic_low_rank}) spans only an $r$-dimensional space. When $n>r$, the joint distribution of ${\tilde{w}}$ is singular. However, we do create a valid stochastic process with covariance function
\begin{equation}\label{eq:low_rank_corr}
\mbox{cov}(\tilde{w}(\ell), \tilde{w}(\ell')) = b_{\theta}^{\T}(\ell)V_{z}b_{\theta}(\ell')\; ,
\end{equation}
where $V_z$ is the variance-covariance matrix (also depends upon parameter $\theta$) for $z$. From (\ref{eq:low_rank_corr}), we see that, even if $b_{\theta}(\cdot,\cdot)$ is stationary, the induced covariance function is not. If the $z$'s are Gaussian, then $\tilde{w}(\ell)$ is a Gaussian process. Every choice of basis functions yields a process and there are too many choices to enumerate here. \cite{wikle_2011} offers an excellent overview of low rank models. 


Different families of spatial models emerge from different specifications for the process $z(\ell)$ and the basis functions $b_{\theta}(\ell,\ell')$. In fact, (\ref{eq:generic_low_rank}) can be used to construct classes of rich and flexible processes. Furthermore, such constructions need not be restricted to low rank models. If dimension reduction is not a concern, then full rank models can be constructed by taking $r=n$ basis functions in (\ref{eq:generic_low_rank}). A very popular specification for $z(\ell)$ is a white noise process so that $z\sim N(0,\sigma^2I_n)$, whereupon (\ref{eq:low_rank_corr}) simplifies to $\sigma^2 b_{\theta}(\ell)^{\T}b_{\theta}(\ell')$. A natural choice for the basis functions is a kernel function, say $b_{\theta}(\ell,\ell') = K_{\theta}(\ell-\ell')$, which puts more weight on $\ell'$ near $\ell$. Variants of this form have been called ``moving average'' models and explored by \cite{barryverhoef1996}, while the term ``kernel convolution'' has been used in a series of papers by Higdon and collaborators \citep[][]{hig98,higswallkern99,hig02} to not only achieve dimension reduction, but also model nonstationary and multivariate spatial processes. The kernel (which induces a parametric covariance function) can depend upon parameters and might even be spatially varying \citep[][]{hig02,paciorek06}. \cite{sanso08} use discrete kernel convolutions of independent processes to construct two different class of computationally efficient spatiotemporal processes.  

Some choices of basis functions can be more computationally efficient than others depending upon the specific application. For example, \cite{cres08} (also see \cite{shicres07}) discuss ``Fixed Rank Kriging'' (FRK) by constructing $B_{\theta}$ using very flexible families of non-stationary covariance functions to carry out high-dimensional kriging, \cite{cres10} extend FRK to spatiotemporal settings calling the procedure ``Fixed Rank Filtering'' (FRF), \cite{katz12} provide efficient constructions for $B_{\theta}$ for massive spatiotemporal datasets, and \cite{katzfuss2013} uses spatial basis functions to capture medium to long range dependence and tapers the residual $w(\ell) - \tildew(\ell)$ to capture fine scale dependence. Multiresolution basis functions \citep[see, e.g.,][]{Nychka_Wikle_Royle_2002,nychka2015} have been shown to be effective in building computationally efficient nonstationary models. These papers amply demonstrate the versatility of low-rank approaches using different basis functions.      

A different approach is to specify the $z(\ell)$ as a spatial process model having a selected covariance function. This process is called the parent process and one can derive a low-rank process $\tilde{w}(\ell)$ from the parent process. For example, one could use the Karhunen-Loeve (infinite) basis expansion for a Gaussian process \citep[see, e.g.,][]{rasm08,ban14} and truncate it to a finite number of terms to obtain a low-rank process. Another example is to project the realizations of the parent process onto a lower-dimensional subspace, which yields the \emph{predictive process} and its variants; see Section~\ref{sec:pred_proc} for details. 

The idea underlying low-rank dimension reduction is not dissimilar to Bayesian linear regression. For example, consider a simplified version of the hierarchical model in (\ref{eq: Bayesian_Spatial_Gaussian_Generic}), where $\beta=0$ and the process parameters $\{\theta, \tau\}$ are fixed. A low rank version of (\ref{eq: Bayesian_Spatial_Gaussian_Generic}) is obtained by replacing $w$ with $B_{\theta}z$, so the joint distribution is
\begin{equation}\label{eq:SWM_Derivation_Normal_Normal}
 N(z\given 0, V_z) \times N(y\given B_{\theta}z, D_{\tau})\; ,
\end{equation}
where $y$ is $n\times 1$, $z$ is $r\times 1$, $D_{\tau}$ and $V_z$ are positive definite matrices of sizes $n\times n$ and $r\times r$, respectively, and $B_{\theta}$ is $n\times r$. The low rank specification is accommodated using $B_{\theta}z$ and the prior on $z$, while $D_{\tau}$ (usually diagonal) has the residual variance components. By computing the marginal covariance matrix $\mbox{var}\{y\}$ in two ways \citep[][]{lindleysmith1972}, one arrives at the well-known Sherman-Woodbury-Morrison formula
\begin{equation}\label{eq: SWM}
 (D_{\tau} + B_{\theta}V_zB_{\theta}^{\T})^{-1} = D_{\tau}^{-1} - D_{\tau}^{-1}B_{\theta}(V_z^{-1} + B_{\theta}^{\T}D_{\tau}^{-1}B_{\theta})^{-1}B_{\theta}^{\T}D_{\tau}^{-1}\; .
\end{equation}   
The above formula reveals dimension reduction in terms of the marginal covariance matrix for $y$. If $D_{\tau}$ is easily invertible (e.g., diagonal), then the inverse of an $n\times n$ covariance matrix of the form $D_{\tau} + B_{\theta}V_zB_{\theta}^{\T}$ can be computed efficiently using the right-hand-side which only involves inverses of $r\times r$ matrices and $D_{\tau}^{-1}$. A companion formula for (\ref{eq: SWM}) is that for the determinant,
\begin{equation}\label{eq: SWM_det}
 \det(D_{\tau} + B_{\theta}V_zB_{\theta}^{\T}) =  \det(V_z)\det(D_{\tau})\det(V_z^{-1} + B_{\theta}^{\T}D_{\tau}^{-1}B_{\theta})\;, 
\end{equation}
which shows that the determinant of the $n\times n$ matrix can be computed as a product of the determinants of two $r\times r$ matrices and that of $D_{\tau}$.

In practical Bayesian computations, however, it is less efficient to directly use the formulas in (\ref{eq: SWM}) and (\ref{eq: SWM_det}). Since both the inverse and the determinant are needed, it is more useful to compute the Cholesky decomposition of the covariance matrix. In fact, one can avoid (\ref{eq: SWM}) completely and resort to a common trick in hierarchical models \citep[see, e.g.,][]{gelman2013} and smoothed ANOVA \citep[][]{hodges2013} that expresses (\ref{eq:SWM_Derivation_Normal_Normal}) as the linear model    
 \begin{align}\label{eq: Normal_Normal_Linear_Model}
 \begin{array}{ccccc}
  \underbrace{\begin{bmatrix} D_{\tau}^{-1/2}y \\ 0\end{bmatrix}} & = & \underbrace{\begin{bmatrix} D_{\tau}^{-1/2} B_{\theta} \\ V_z^{-1/2} \end{bmatrix}}{z} & + & \underbrace{\begin{bmatrix} e_1 \\ e_2\end{bmatrix}} \\
  y_{\ast} &  & B_{\ast} & & e_{\ast} 
 \end{array}\;,\; \mbox{ where }\; e_{\ast} \sim N(0, I_{n+r})\;,
 \end{align}
$V_z^{1/2}$ and $D_{\tau}^{1/2}$ are matrix square roots of of $V_z$ and $D_{\tau}$, respectively. For example, in practice $D_{\tau}$ is diagonal so $D_{\tau}^{1/2}$ is simply the square root of the diagonal elements of $D_{\tau}$, while $V_z^{1/2}$ is the triangular (upper or lower) Cholesky factor of the $r\times r$ matrix $V_{z}$. The marginal density of $p(y_{\ast}\given \theta,\tau)$ after integrating out $z$ now corresponds to the linear model $y_{\ast} = B_{\ast}\hat{z} + e_{\ast}$, where $\hat{z}$ is the ordinary least-square estimate of $z$. Such computations are easily conducted in statistical programming environments such as \texttt{R} by applying the \texttt{chol} function to obtain the Cholesky factor $V_z^{1/2}$, a \texttt{backsolve} function to efficiently obtain $V_z^{-1/2}z$ in constructing (\ref{eq: Normal_Normal_Linear_Model}), and an \texttt{lm} function to compute the least squares estimate of $z$ using the QR decomposition of the design matrix $B_{\ast}$. We discuss implementation of low rank hierarchical models in a more general contexts in Section~\ref{sec: low_rank_implementation}. 

\subsection{Biases in low-rank models}\label{sec:low_rank_biases}

\noindent Irrespective of the precise specifications, low-rank models tend to underestimate uncertainty (since they are driven by a finite number of random variables), hence, overestimate the residual variance (i.e., the nugget). Put differently, this arises from systemic over-smoothing or model under-specification by the low-rank model when compared to the parent model. For example, if $w(\ell) = \tildew(\ell) + \eta(\ell)$, where $w(\ell)$ is the parent process and $\tildew(\ell)$ is a low-rank approximation, then ignoring the residual $\eta(\ell) = w(\ell)-\tildew(\ell)$ can result in loss of uncertainty and oversmoothing. In settings where the spatial signal is weak compared to the noise, such biases will be less pronounced. Also, it is conceivable that in certain specific case studies proper choices of basis functions (e.g., multiresolution basis functions) will be able to capture much of the spatial behavior and the effect of the bias will be mitigated. However, in general it will be preferable to develop models that will be able to compensate for the overestimation of the nugget.    

This phenomenon, in fact, is not dissimilar to what is seen in linear regression models and is especially transparent from writing the parent likelihood and low-rank likelihood as mixed linear models. To elucidate, suppose, without much loss of generality, that ${\cal U}$ is a set with $n$ points of which the first $r$ act as the knots. Let us write the Gaussian likelihood with the parent process as $N(y\given Bu, \tau^2I)$, where $B$ is the $n\times n$ lower-triangular Cholesky factor of $K_{\theta}$ ($B = B_{\theta}$ depends on $\theta$, but we suppress this here) and $u = (u_1,u_2,\ldots,u_n)^{\T}$ is now an $n\times 1$ vector such that $u_i \stackrel{iid}{\sim} N(0,1)$. Writing $B = [B_{1} : B_{2}]$, where $B_{1}$ has $r < n$ columns, suppose we derive a low-rank model by truncating to only the first $r$ basis functions. The corresponding likelihood is $N(y\given B_{1}\tilde{u}_1, \tau^2 I)$, where $\tilde{u}_1$ is an $r\times 1$ vector whose components are independently and identically distributed $N(0,1)$ variables. Customary linear model calculations reveal that the magnitude of the residual vector from the parent model is given by $y^{\T}(I - P_{B})y$, while that from the low-rank model is given by $y^{\T}(I - P_{B_1})y$, where $P_{A}$ denotes the orthogonal projector matrix onto the column space of any matrix $A$. Using the fact that $P_{B} = P_{B_1} + P_{[(I-P_{B_1})B_2]}$, which is a standard result in linear model theory, we find the excess residual variability in the low-rank likelihood is summarized by $y^{\T}P_{[(I-P_{B_1})B_2]}y$ which can be substantial when $r$ is much smaller than $n$.

\begin{figure}[t!]
\begin{center}
\includegraphics[width=3in,height=3in]{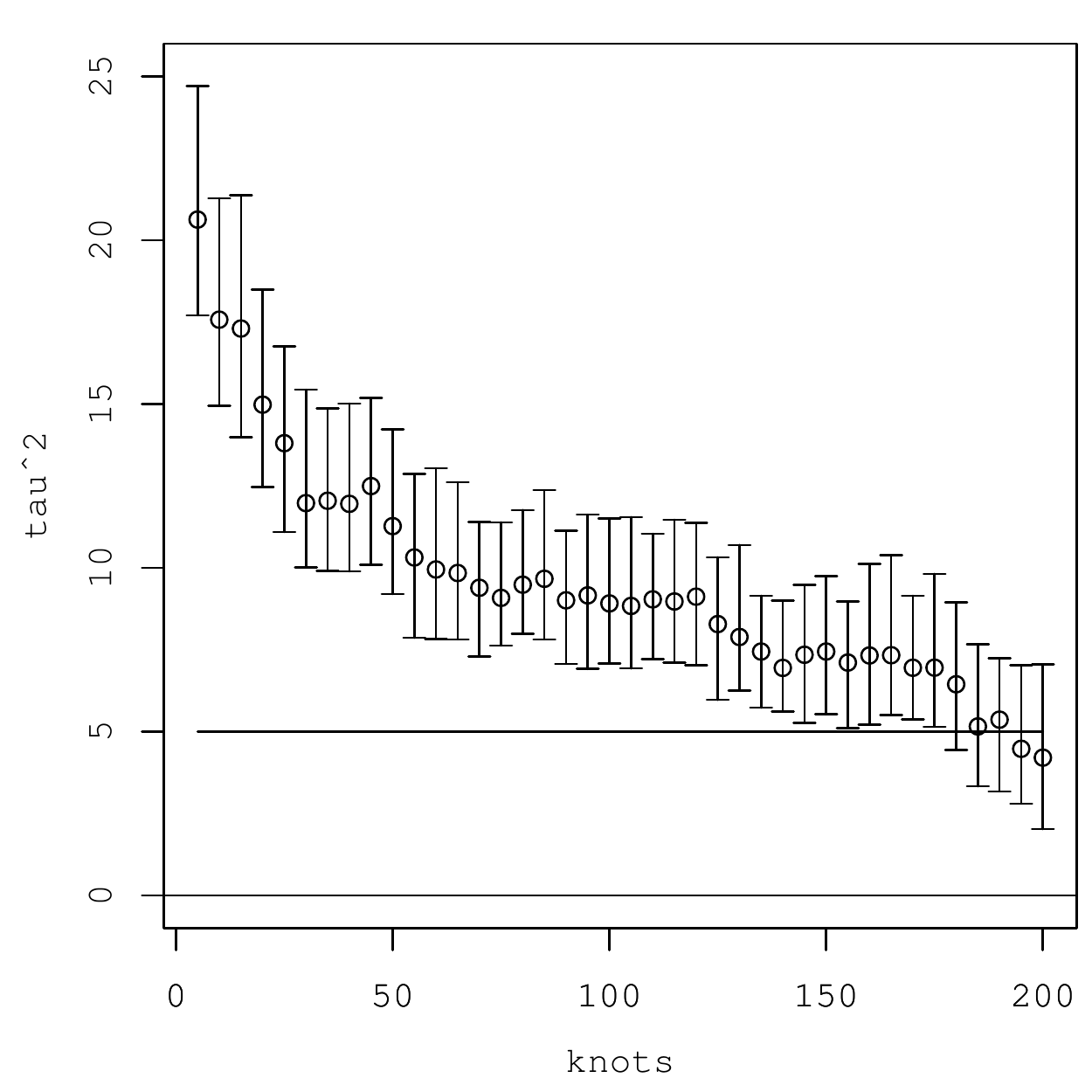}
\end{center}
\vspace{-0.5cm}
\caption{95\% credible intervals for the nugget for 40 different low-rank radial-basis models with knots varying between $5$ and $200$ in steps of $5$. The horizontal line at $\tau^2=5$ denotes the true value of $\tau^2$ with which the data was simulated.}\label{fig: low_rank_nugget}
\end{figure}

In practical data analysis, the above phenomenon is usually manifested by an overestimation of the nugget variance as it absorbs the residual variation from the low-rank approximation. Consider the following simple experiment. We simulated a spatial dataset using the spatial regression model in (\ref{eq: BasicModel}) with $n=200$ fixed spatial locations, say $\{\ell_1,\ell_2,\ldots,\ell_n\}$, within the unit square, and setting $\beta=0$, $\tau^2=5$, $w(\ell)\sim GP(0, K_{\theta})$, where $K_{\theta}(\ell_i,\ell_j) = \sigma^2\exp(-\phi\|\ell_i-\ell_j\|)$ with $\sigma^2=5$ and $\phi=9$. We then fit the low rank model (\ref{eq:SWM_Derivation_Normal_Normal}) with $D=\tau^2I_{n\times n}$, $V=I_{r\times r}$, and $B$ as the $n\times r$ matrix with $i$-th row $b^{\T}(\ell_i) = K_{\theta}(\ell_i,\calU^*) K^{-1/2}_{\theta}(\calU^*,\calU^*)$, where $\calU^*=\{\ell^*_1,\ell_2^*,\ldots,\ell_r^*\}$ is a set of $r$ knots, $K_{\theta}(\ell_i,\calU^*)$ is the $1\times r$ vector with $j$-th element $K_{\theta}(\ell_i,\ell_j^*)$ and $K^{-1/2}_{\theta}(\calU^*,\calU^*)$ is the inverse of the lower-triangular Cholesky factor of the $r\times r$ matrix with elements $K_{\theta}(\ell_i^*,\ell_j^*)$. This emerges from using low-rank radial basis functions in (\ref{eq:generic_low_rank}); \citep[see, e.g., ][]{ruppwandcarroll2003}. We fit $40$ such models increasing $r$ from $5$ to $200$ in steps of $5$. Figure~\ref{fig: low_rank_nugget} presents the 95\% posterior credible intervals for $\tau^2$. Even with $r=175$ knots for a dataset with just $200$ spatial locations, the estimate of the nugget was significantly different from the true value of the parameter. This indicates that low rank processes may be unable to accurately estimate the nugget from the true process. Also, they will likely produce oversmoothed interpolated maps of the underlying spatial process and impair predictive performance. As one specific example, Table~4 in \cite{ban08} report less than optimal posterior predictive coverage from a predictive process model (see Section~\ref{sec:pred_proc}) with over 500 knots for a dataset comprising 15,000 locations.   

Although this excess residual variability can be quantified as above (for any given value of the covariance parameters $\theta$), it is less clear how the low-rank likelihood could be modified to compensate for this oversmoothing without adding significantly to the computational burden. Matters are complicated by the fact that expressions for the excess variability will involve the unknown process parameters $\theta$, which must be estimated. In fact, not all low-rank models deliver a straightforward quantification for this bias. For instance, low-rank models based upon kernel convolutions approximate $w(\ell)$ with $w_{KC}(\ell)=\sum_{j=1}^{n^{\ast}}K_{\theta}(\ell-\ell_j^{\ast},\theta)u_j$, where $K_{\theta}(\cdot)$ is some kernel function and $u_j\stackrel{iid}{\sim} N(0,1)$, assumed to arise from a Brownian motion $U(\omega)$ on $\Re^2$. The difference $w(\ell)-w_{KC}(\ell)$ does not, in general, render a closed form and may be difficult to approximate efficiently.

\subsection{Predictive process models and  variants}\label{sec:pred_proc}
\noindent One particular class of low-rank processes have been especially useful in providing easy tractability to the residual process. Let $w(\ell) \sim GP(0, K_{\theta}(\cdot,\cdot))$ and let $w^*$ be the $r\times 1$ vector of $w(\ell_j^*)$'s over a set $\calU^*$ of $r$ knots. The usual spatial interpolant (that leads to ``kriging'') at an arbitrary site $\ell$ is  
\begin{equation}\label{eq: predictive_process}
\tildew(\ell)= \mbox{E}[w(\ell)\given w^{\ast}] = K_{\theta}(\ell,\calU^*)K_{\theta}^{-1}(\calU^*, \calU^*)w^{\ast}\; .
\end{equation}
This single site interpolator, in fact, is a well-defined process $\tildew(\ell)\sim GP(0, \tilde{K}_{\theta}(\cdot, \cdot))$ with covariance function, $\tilde{K}_{\theta}(\ell,\ell') = K_{\theta}(\ell;\calU^*)K^{-1}_{\theta}(\calU^*,\calU^*)K_{\theta}(\calU^*,\ell')$\;.
We refer to $\tildew(\ell)$ as the \emph{predictive process} derived from the \emph{parent process} $w(\ell)$. The realizations of $\tilde{w}(\ell)$ are precisely the kriged predictions conditional upon a realization of $w(\ell)$ over ${\cal U}^{\ast}$. The process is completely specified given the covariance function of the parent process and the set of knots, ${\cal U}^{\ast}$. The corresponding basis functions in (\ref{eq:generic_low_rank}) are given by $b^{\T}_{\theta}(\ell) = K_{\theta}(\ell, \calU^*)K_{\theta}^{-1}(\calU^*, \calU^*)$. These methods have are referred to as subset of regressors in Gaussian process regressions for large data sets in machine learning \citep[][]{quinonerorasmussen05,rasm08}. \cite{ban08} coined the term predictive process (as the process could be derived from kriging equations) and developed classes of scalable Bayesian hierarchical spatial process models by replacing the parent process with its predictive process counterpart. An alternate derivation is available by truncating the Karhunen-Loeve (infinite) basis expansion for a Gaussian process to a finite number of terms and solving (approximately) the integral eigen-system equation for $K_{\theta}(\ell,\ell')$ by an approximate linear system over the set of knots \citep[see, e.g.,][]{rasm08,sang12,ban14}.

Exploiting elementary properties of conditional expectations, we obtain
\begin{equation}\label{eq: predictive_process_bias_inequality}
 \mbox{var}\{w(\ell)\} = \mbox{var}\{\mbox{E}[w(\ell)\given w^*]\} + \mbox{E}\{\mbox{var}[w(\ell)\given w^*]\} \geq \mbox{var}\{\mbox{E}[w(\ell)\given w^*]\}\; ,
\end{equation}
which implies that $\mbox{var}\{w(\ell)\} \geq \mbox{var}\{\tildew(\ell)\}$ and the variance of $\eta(\ell) = w(\ell) - \tildew(\ell)$ is simply the difference of the variances. For Gaussian processes, we get the following closed form for $\mbox{Cov}\{\eta(\ell), \eta(\ell')\}$,
\begin{equation}\label{eq: low_rank_residual_covariance}
 K_{\eta, \theta}(\ell,\ell') = K_{\theta}(\ell,\ell') - K_{\theta}(\ell, \calU^*)K_{\theta}^{-1}(\calU^*,\calU^*)K_{\theta}(\calU^*,\ell')\; .
\end{equation}
Therefore, $\mbox{var}\{\eta(\ell)\} =  K_{\eta, \theta}(\ell,\ell)$, which we denote as $\delta^2(\ell)$.

Perhaps the simplest way to remedy the bias in the predictive process is to approximate the residual process $\eta(\ell)$ with a heteroskedastic process $\tilde{\epsilon}(\ell)\stackrel{ind}{\sim} N(0,\delta^2(\ell))$. We construct a \emph{modified} or \emph{bias-adjusted} predictive process as
\begin{equation}\label{wq: modified_predictive_process}
 \tildew_{\epsilon}(\ell) = \tildew(\ell) + \tilde{\epsilon}(\ell)\; ,
\end{equation}
where $\tilde{\epsilon}(\ell)$ is independent of $\tildew(\ell)$. It is easy to see that $\mbox{var}\{\tildew_{\epsilon}(\ell)\} = \mbox{var}\{w(\ell)\}$, so the variance of the two processes are the same. Also, the remedy is computationally efficient -- adding an independent space-varying nugget does not incur substantial computational expense. \cite{Finley_etall_2009} offer computational details for the modified predictive process, while \cite{ban10} show the effectiveness of the bias adjustment in mitigating the effect exhibited in Figure~\ref{fig: low_rank_nugget} and in estimating multiple variance components in the presence of different structured random effects.

We present a brief simulation example revealing the benefits of the modified predictive process. We generate 2000 locations within a $[0,100]\times[0,100]$ square and then generate the outcomes from (\ref{eq: BasicModel}) using only an intercept as the regressor, an exponential covariance function with range parameter $\phi=0.06$ (i.e., such that the spatial correlation is $\sim 0.05$ at 50 distance units), scale $\sigma^2=1$ for the  spatial process, and with  nugget variance $\tau^2=1$. We then fit the predictive process and modified predictive process models derived from (\ref{eq: BasicModel}) using a hold out set of randomly selected sites, along with a separate set of regular lattices for the knots ($m=49$, $144$ and $900$). Table~\ref{UnivSimIll} shows the posterior estimates and the square roots of MSPE based on the predictions for the hold-out data. The overestimation of $\tau^2$ by the predictive process is apparent and we also see how the modified predictive process is able to adjust for the $\tau^2$. Not surprisingly, the RMSPE is essentially the same under either process model.
{\small
\begin{table}[t]
\begin{center}
{\small
\begin{tabular}{ccccc}
\hline
   & $\mu$ & $\sigma^2$ & $\tau^2$ & RMSPE  \\
\hline
\hline
True & 1 & 1 & 1 &  \\
$m=49$ & & & & \\
PP & 1.37 (0.29,2.61)& 1.37 (0.65,2.37)& 1.18 (1.07,1.23)& 1.21\\
MPP & 1.36 (0.51,2.39)& 1.04 (0.52,1.92) & 0.94 (0.68.1,14) & 1.20\\
$m=144$ & & & & \\
PP & 1.36 (0.52,2.32)& 1.39 (0.76,2.44)& 1.09 (0.96, 1.24) & 1.17\\
MPP & 1.33 (0.50,2.24)& 1.14 (0.64,1.78)& 0.93 (0.76,1.22) & 1.17 \\
$m=900$ & & & & \\
PP & 1.31 (0.23, 2.55) & 1.12 (0.85,1.58)& 0.99 (0.85,1.16) &  1.17\\
MPP & 1.31 (0.23,2.63)& 1.04 (0.76,1.49) & 0.98 (0.87,1.21) & 1.17 \\
\hline
\end{tabular}
}
\end{center}
\caption{Parameter estimates for the predictive process (PP) and modified predictive process (MPP) models in the univariate simulation.}\label{UnivSimIll}
\end{table}
}

Further enhancements to the modified predictive process are possible. Since the modified predictive process adjusts only the variance, information in the covariance induced by the residual process $\eta(\ell)$ is lost. One alternative is to use the so called ``full scale approximation'' proposed by \cite{sang11} and \cite{sang12}, where $\eta(\ell)$ is approximated by a tapered process, say $\eta_{tap}(\ell)$. The covariance function for $\eta(\ell)$ is of the form $K_{\eta,\theta}(\ell, \ell')K_{tap,\nu}(\|\ell-\ell'\|)$, where $K_{\eta,\theta}(\ell, \ell')$ is as in (\ref{eq: low_rank_residual_covariance}) and $K_{tap,\nu}(\|\ell-\ell'\|)$ is a compactly supported covariance function that equals $0$ beyond a distance $\nu$ \citep[see, e.g.,][ for some practical choices.]{fur06}. This full scale approximation is also able to more effectively capture small scale dependence. \cite{katzfuss2013} extended some of these ideas by modeling the spatial error as a combination of a low-rank component designed to capture medium to long-range dependence and a tapered component to capture local dependence. 

Perhaps the most promising use of the predictive process, at least in terms of scalability to massive spatial datasets, is the recent multiresolution approximation proposed by \cite{katzfussmultires}. Instead of approximating the residual process $\eta(\ell)$ in one step, the idea here is to partition the spatial domain recursively and construct a sequence of approximations. We start by partitioning the domain of interest $\calL$ into $J$ non-intersecting subregions, say $\calL_1,\calL_2,\ldots,\calL_J$, such that $\calL = \cup_{j=1}^J\calL_j$. We call the $\calL_j$'s level-1 subregions. We fix a set of knots in $\calL$ and write the parent process as $w(\ell) = \tildew(\ell) + \eta(\ell)$, where $\tildew(\ell)$ is the predictive process as in (\ref{eq: predictive_process}) and $\eta(\ell)$ is the residual Gaussian process with covariance function given by (\ref{eq: low_rank_residual_covariance}). At resolution 1, we replace $\eta(\ell)$ with a block-independent process $\eta_1(\ell)$ such that $\mbox{Cov}\{\eta_1(\ell), \eta_1(\ell')\} = 0$ if $\ell$ and $\ell'$ are not in the same subregion and is equal to (\ref{eq: low_rank_residual_covariance}) if $\ell$ and $\ell'$ are in the same subregion. 

At the second resolution, each $\calL_j$ is partitioned into a set of disjoint subregions $\calL_{j1},\calL_{j2},\ldots,\calL_{jm}$. We call these the level-2 subregions and choose a set of knots within each. We approximate $\eta_1(\ell) \approx \tilde{\eta_1}(\ell) + \eta_2(\ell)$, where $\tilde{\eta_1}(\ell)$ is the predictive process derived from $\eta_1(\ell)$ using the knots in $\calL_j$ if $\ell \in \calL_j$ and $\eta_2(\ell)$ is the analogous block-independent approximation across the subregions within each $\calL_j$. Thus, $\mbox{Cov}\{\eta_2(\ell),\eta_2(\ell')\} = 0$ if $\ell$ and $\ell'$ are not in the same level-2 subregion and will equal $\mbox{Cov}\{\eta_1(\ell),\eta_1(\ell')\}$ when $\ell$ and $\ell'$ are in the same level-2 subregion. At resolution 3 we partition each of the level-2 subregions into level-3 subregions and continue the approximation of the residual process from the predictive process. At the end of $M$ resolutions, we arrive at the mult-resolution predictive process $\tildew_{M}(\ell) = \tildew(\ell) + \sum_{i=1}^{M-1}\tilde{\eta}_i(\ell) + \eta_{M}(\ell)$, which, by construction, is a valid Gaussian process. The computational complexity with the multi-resolution predictive process is $\sim O(nM^2r^2)$, where $M$ is the number of resolutions and $r$ is the number of knots chosen within each subregion.    
 
To summarize, we do not recommend the use of \emph{just} a reduced/low rank model. To improve performance, it is necessary to approximate the residual process and, in this regard, the predictive process is especially attractive since the residual process is available explicitly.  
  
\subsection{Bayesian implementation for low-rank models}\label{sec: low_rank_implementation}
\noindent A very rich and flexible class of spatial and spatiotemporal models emerge from the hierarchical linear mixed model
\begin{align}\label{eq: Bayesian_Spatial_Gaussian_Generic_Implementation}
p(\theta) \times p(\tau) \times N(\beta\given \mu_{\beta}, V_{\beta})\times N(z\given 0, V_{z,\theta}) \times N(y \given X\beta + B_{\theta}z, D_{\tau})\; , 
\end{align}
where $y$ is an $n\times 1$ vector of possibly irregularly located observations, $X$ is a known $n\times p$ matrix of regressors ($p < n$), $V_{u,\theta}$ and $D_{\tau}$ are families of $r\times r$ and $n\times n$ covariance matrices depending on unknown process parameters $\theta$ and $\tau$, respectively, and $B_{\theta}$ is $n\times r$ with $r\leq n$. The low-rank models in (\ref{eq:generic_low_rank}) emerge when $r << n$ and $B_{\theta}$ is the matrix obtained by evaluating the basis functions. Proper prior distributions $p(\theta)$ and $p(\tau)$ for $\theta$ and $\tau$, respectively, complete the hierarchical specification. 

Bayesian inference proceeds, customarily, by sampling $\{\beta,z,\theta, \tau\}$ from (\ref{eq: Bayesian_Spatial_Gaussian_Generic_Implementation}) using Markov chain Monte Carlo (MCMC) methods. For faster convergence, we integrate out $z$ from the model and first sample from $\displaystyle p(\theta, \tau, \beta \given y) \propto p(\theta)\times p(\tau) \times N(\beta\given \mu_{\beta}, V_{\beta})\times N(y\given X\beta, \Sigma_{y\given\theta,\tau})$,
where $\Sigma_{y\given\theta,\tau} = B_{\theta}V_{z,\theta}B_{\theta}^\top + D_{\tau}$. 
Working directly with $\Sigma_{y\given\theta,\tau}$ will be expensive. Usually $D_{\tau}$ is diagonal or sparse, so the expense is incurred from the matrix $B_{\theta}V_{z,\theta}B_{\theta}^\top$. Assuming that $B_{\theta}$ and $V_{z,u}$ are computationally inexpensive to construct for each $\theta$ and $\tau$,  $B_{\theta}V_{z,\theta}B_{\theta}^\top$ requires $\sim O(rn^2)$ flops. Using the Sherman-Woodbury-Morrison formula in (\ref{eq: SWM}) will avoid constructing $B_{\theta}V_{z,\theta}B_{\theta}^\top$ or inverting any $n\times n$ matrix. However, in practice it is better to not directly compute the right hand side of (\ref{eq: SWM}) as it involves some redundant matrix multiplications. Furthermore, we wish to obtain the determinant of $\Sigma_{y\given\theta,\tau}$ cheaply. These are efficiently accomplished as outlined below.

The primary computational bottleneck lies in evaluating the multivariate Gaussian likelihood $N(y\given X\beta, \Sigma_{y\given\theta,\tau})$ which is required for updating the parameters $\{\theta, \tau\}$ (e.g., using random-walk Metropolis or Hamiltonian Monte Carlo steps). We can accomplish this effectively using two functions: $L = \texttt{chol}(V)$ which computes the Cholesky factorization for any positive definite matrix $V = LL^{\top}$, where $L$ is lower-triangular, and $W = \texttt{trsolve}(T,B)$ which solves the triangular system $TW=B$ for a triangular (lower or upper) matrix $T$. We first compute    
\begin{align}\label{eq: SWM_Compute}
 \left(B_{\theta}V_{z,\theta}B_{\theta}^{\top} + D_{\tau}\right)^{-1} &= 
 D_{\tau}^{-1/2}(I - H^{\top}H)D_{\tau}^{-1/2}\;,
\end{align}
where $H$ is obtained by first computing $W=D^{-1/2}B_{\theta}$, then the Cholesky factorization $L = \texttt{chol}(V_{z,\theta}^{-1} + W^{\top}W)$, and finally solve the triangular system $H = \texttt{trsolve}(L,W^{\top})$.  Having obtained $H$, we compute $e=y-X\beta$, $m_1 = D^{-1/2}e$, $m_2 = Hm_1$, and obtain $T = \texttt{chol}(I_{r}-HH^{\top})$. The log-target density for $\{\theta, \tau\}$ is then computed as
\begin{align}\label{eq: Metrop_target_density}
& \log p(\theta) + \log p(\tau) - \frac{1}{2}\sum_{i=1}^{n} d_{ii} + \sum_{i=1}^{r} \log t_{ii} - \frac{1}{2}(m_1^{\top}m - m_2^{\top}m_2)\; ,
\end{align}
where $d_{ii}$'s and $t_{ii}$'s are the diagonal elements of $D_{\tau}$ and $T$, respectively. The total number of flops required for evaluating the target is $O(nr^{2} + r^3) \approx O(nr^2)$ (since $r << n$) which is considerably cheaper than the $O(n^3)$ flops that would have been required for the analogous computations in a full Gaussian process model. In practice, Gaussian proposal distributions are employed for the Metropolis algorithm and all parameters with positive support are transformed to their logarithmic scale. Therefore, the necessary Jacobian adjustments are made to (\ref{eq: Metrop_target_density}) by adding some scalar quantities with negligible computational costs.

Starting with initial values for all parameters, each iteration of the MCMC executes the above calculations to provide a sample for $\{\theta,\tau\}$. The regression parameter $\beta$ is then sampled from its full conditional distribution. Writing $\Sigma_y = B_{\theta}V_{z,\theta}B_{\theta}^{\T} + D_{\tau}$ as in (\ref{eq: SWM_Compute}), the full conditional distribution for $\beta$ is $N(Aa, A)$, where $A^{-1} = \Sigma_{\beta}^{-1} + X^{\top}\Sigma_y^{-1}X$ and $a = \Sigma_{\beta}^{-1}\mu_{\beta} + X^{\T}\Sigma_y^{-1}y$. These are efficiently computed as $[f : F] = D^{-1/2}[y : X]$, $\tilde{F} = HF$ and setting $a = \Sigma_{\beta}^{-1}\mu_{\beta} + F^{\T}f - \tilde{F}^{\T}Hf$ and $L = \texttt{chol}(\Sigma_{\beta}^{-1} + F^{\T}F-\tilde{F}^{\T}\tilde{F})$. We then compute $\beta = \texttt{trsolve}(L^{\T}, \texttt{trsolve}(L,a)) + \texttt{trsolve}(L,\tilde{Z})$, where $\tilde{Z}$ is a conformable vector of independent $N(0,1)$ variables. 

We repeat the above computations for each iteration of the MCMC algorithm using the current values of the process parameters in $\Sigma_y$. The algorithm described above will produce, after convergence, posterior samples for $\Omega = \{\theta,\tau,\beta\}$. We then sample from the posterior distribution $\displaystyle p(z\given y) = \int p(z\given \Omega, y) p(\Omega\given y)d\Omega$, where $p(z\given \Omega, y) = N(z\given Aa, A)$ with $A = (V_{z,\theta}^{-1} + B_{\theta}^{\T}D_{\tau}^{-1}B_{\theta})^{-1}$ and $a = B_{\theta}^{\top}D_{\tau}^{-1}(y - X\beta)$. For each $\Omega$ drawn from $p(\Omega\given y)$ we will need to draw a corresponding $z$ from $N(z\given Aa, A)$. This will involve $\texttt{chol}(A)$. Since the number of knots $r$ is usually fixed at a value much smaller than $n$, obtaining $\texttt{chol}(A)$ is $\sim O(r^3)$ and not as expensive. However, it will involve the inverse of $V_{z,\theta}$, which is computed using $\texttt{chol}(V_{z,\theta})$ and can be numerically unstable for certain smoother covariance functions such as the Gaussian or the Mat\'ern with large $\nu$. A numerically more stable algorithm exploits the relation $A = Q - Q(V_{z,\theta} + Q)^{-1}Q$, where $Q^{-1} = B_{\theta}^{\top}D_{\tau}^{-1}B_{\theta}$. 
For each $\Omega$ sampled from $p(\Omega\given y)$, we compute $L = \texttt{chol}(V_{z,\theta} + Q)$, $W = \texttt{trsolve}(L,Q)$ and $L = Q - W^{\top}W$. We generate an $r\times 1$ vector $Z^*\sim N(0, I_{r})$ and set $z = L(Z^* + L^{\top}a)$. 
Repeating this for each $\Omega$ drawn from $p(\Omega\given y)$ produces a sample of $z$'s from $p(z\given y)$. 

Finally, we seek predictive inference for $y(\ell_0)$ at any arbitrary space-time coordinate $\ell_0$. Given $x^{\top}(\ell_0)$, we draw $y(\ell_0) \sim N\left(x^{\top}(\ell_0)\beta + b_{\theta}^{\T}(\ell_0)z, \tau^2\right)$ for every posterior sample of $\Omega$ and $z$. This yields the corresponding posterior predictive samples for $z(\ell_0)$ and $y(\ell_0)$. Posterior predictive samples of the latent processes can also be easily computed as $z(\ell_0) = b_{\theta}^{\T}(\ell_0)z$ for each posterior sample of the $z$ and $\theta$. Posterior predictive distributions at any of the observed $\ell_i$'s yield \emph{replicated} data \citep[see, e.g.,][]{gelman2013} that can be used for model assessment and comparisons. \cite{finbangel15} provide more extensive implementation details for models such as (\ref{eq: Bayesian_Spatial_Gaussian_Generic_Implementation}) in the context of the \texttt{spBayes} package in \texttt{R}.   

\section{Sparsity-inducing nearest-neighbor Gaussian processes}\label{sec:sparsity}
\noindent Low-rank models have been, and continue to be, widely employed for analyzing spatial and spatiotemporal data. The algorithmic cost for fitting low-rank models typically decrease from $O(n^3)$ to $O(nr^2 + r^3 ) \approx O(nr^2)$ flops since $n >> r$. However, when $n$ is large, empirical investigations suggest that $r$ must be fairly large to adequately approximate the parent process and the $nr^2$ flops become exorbitant. Furthermore, low-rank models can perform poorly depending upon the smoothness of the underlying process or when neighboring observations are strongly correlated and the spatial signal dominates the noise \citep[][]{stein2014}. 

\begin{figure}[t]
\begin{center}
\subfigure[True w]{\includegraphics[width=1.75in]{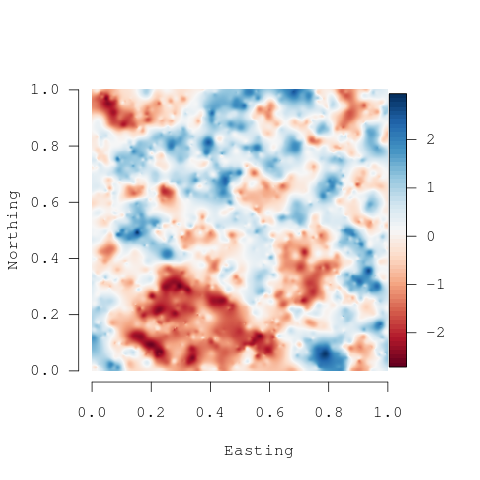}\label{uni-w-obs}}
\subfigure[Full GP]{\includegraphics[width=1.75in]{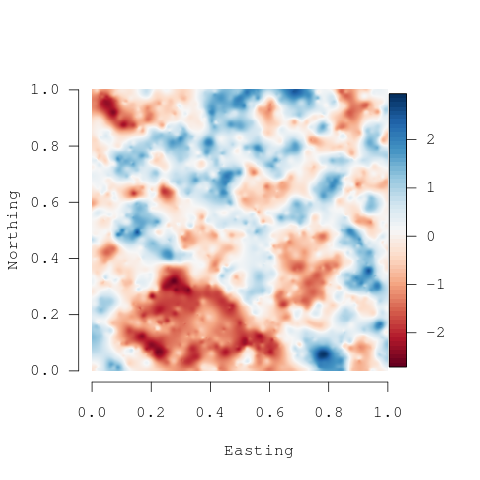}\label{uni-w-gs}}
\subfigure[PPGP 64 knots]{\includegraphics[width=1.75in]{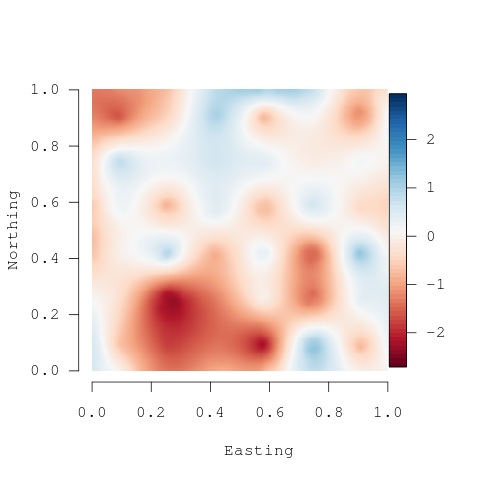}\label{uni-pp64-gs}}\\
\end{center}
\caption{Comparing estimates of a simulated random field using a full Gaussian Process (Full GP) and a Gaussian Predictive process (PPGP) with 64 knots. The oversmoothing by the low-rank predictive process is evident.} \label{fig:uni-w}
\end{figure}

As an example, consider part of the simulation experiment presented in \cite{datta16}, where a spatial random field was generated over a unit square using a Gaussian process with fixed spatial process parameters over a set of $2500$ locations. We then fit a full Gaussian process model and a predictive process model with $64$ knots. Figure~\ref{fig:uni-w} presents the results \citep[see, e.g.,][ for details.]{datta16} While the estimated random field from the full Gaussian process is almost indistinguishable from the true random field, the surface obtained from the predictive process with $64$ locations substantially oversmooths. This oversmoothing can be ameliorated by using a larger number of knots, but that adds to the computational burden.   

Figure~\ref{fig:uni-w} serves to reinforce findings that low-rank models may be limited in their ability to produce accurate representation of the underlying process at massive scales. They will need a considerably larger number of basis functions to capture the features of the process and will require substantial computational resources for emulating results from a full GP. As the demands for analyzing large spatial datasets increase from the order of $\sim 10^4$ to $\sim 10^6$ locations, low-rank models may struggle to deliver acceptable inference. In this regard, enhancements such as the multi-resolution predictive process approximations referred to in Section~\ref{sec:pred_proc} are highly promising. 

An alternative is to develop full rank models that can exploit sparsity. 
Instead of deriving basis approximations for $w$, one could achieve computational gains by modeling either its covariance function or its inverse as sparse. Covariance tapering does the former by modeling $\mbox{var}\{w\} = K_{\theta} \odot K_{\mbox{tap},\nu}$, where $K_{\mbox{tap},\nu}$ is a sparse covariance matrix formed from a compactly supported, or \emph{tapered}, covariance function with tapering parameter $\nu$ and $\odot$ denotes the element wise (or Hadamard) product of two matrices. The Hadamard product of two positive definite matrices is again a positive definite matrix, so $K_{\theta} \odot K_{\mbox{tap},\nu}$ is positive definite. Furthermore, $K_{\mbox{tap},\nu}$ is sparse because a tapered covariance function is equal to $0$ for all pairs of locations separated by a distance beyond a threshold $\nu$. We refer the reader to \cite{fur06}, \cite{kauf08} and \cite{du09} for further computational and theoretical details on covariance tapering. Covariance tapering is undoubtedly an attractive approach for constructing sparse covariance matrices, but its practical implementation for full Bayesian inference will generally require efficient sparse Cholesky decompositions, numerically stable determinant computations and, perhaps most importantly, effective memory management. These issues are yet to be tested for truly massive spatiotemporal datasets with $n \sim 10^5$ or more.     

Another way to exploit sparsity is to model the inverse of $\mbox{var}\{w\}$ as a sparse matrix. For finite-dimensional distributions conditional and simultaneous autoregressive (CAR and SAR) models \citep[see, e.g.,][and references therein]{cres93,ban14} adopt this approach for areally referenced datasets. More generally, Gaussian Markov random fields or GMRFs \citep[see, e.g.,][]{rueheld04} are widely used tools for constructing sparse precision matrices and have led to computational algorithms such as the Integrated Nested Laplace Approximation (INLA) developed by \cite{ruemartinochopin2009}. A subsequent article by \cite{lindgrenruelindstrom2011} show how Gaussian processes can be approximated by GMRFs using computationally efficient sparse representations. Thus, a Gaussian process model with a dense covariance function is approximated by a GMRF with a sparse precision matrix. The approach is very computationally efficient for certain classes of covariance functions generated by a certain class of stochastic partial differential equations (including the versatile Mat\'ern class), but their inferential performance on unobservable spatial, spatiotemporal or multivariate Gaussian processes (perhaps specified through more general covariance or cross-covariance functions) embedded within Bayesian hierarchical models is yet to be assessed.
 
Rather than working with approximations to the process, one could also construct massively scalable sparsity-inducing Gaussian processes that can be conveniently embedded within Bayesian hierarchical models and deliver full Bayesian inference for random fields at arbitrary resolutions. Section~\ref{sec: sparse_ggm} describes how sparsity is introduced in the precision matrices for graphical Gaussian models by exploiting the relationship between the Cholesky decomposition of a positive definite matrix and conditional independence. These sparse Gaussian models (i.e., normal distributions with sparse precision matrices) can be used prior models for a finite number of spatial random effects. Section~\ref{sec: nngp} shows the construction of a process from these graphical Gaussian models. This process will be a Gaussian process whose finite-dimensional realizations will have sparse precision matrices. We call them Nearest Neighbor Gaussian Processes (NNGP). Finally, Section~\ref{sec: hierarchical_nngp_models} outlines how the process can be embedded within hierarchical models and presents some brief simulation examples demonstrating certain aspects of inference from NNGP models. 

\subsection{Sparse Gaussian graphical models}\label{sec: sparse_ggm}

\noindent Consider the hierarchical model (\ref{eq: Bayesian_Spatial_Gaussian_Generic}) and, in particular, the expensive prior density $N(w\given 0, K_{\theta})$. From the dense covariance matrix $K_{\theta}$, we wish to obtain a covariance matrix $\tildeK_{\theta}$ such that $\tildeK_{\theta}^{-1}$ is sparse and, importantly, its determinant is available cheaply. What would be an effective way of achieving this? One approach would be to consider \emph{modeling} the Cholesky decomposition of the precision matrix so that it is sparse. For example, forcing some elements in the dense half of the triangular Cholesky factor to be zero will introduce sparsity in the precision matrix. To precisely set out which elements should be made zero in the Cholesky factor, we borrow some fundamental notions of sparsity from graphical (Gaussian) models.

\begin{figure}[t]
\begin{center}
\subfigure[Full graph]{\includegraphics[width=2.5in,height=2.5in]{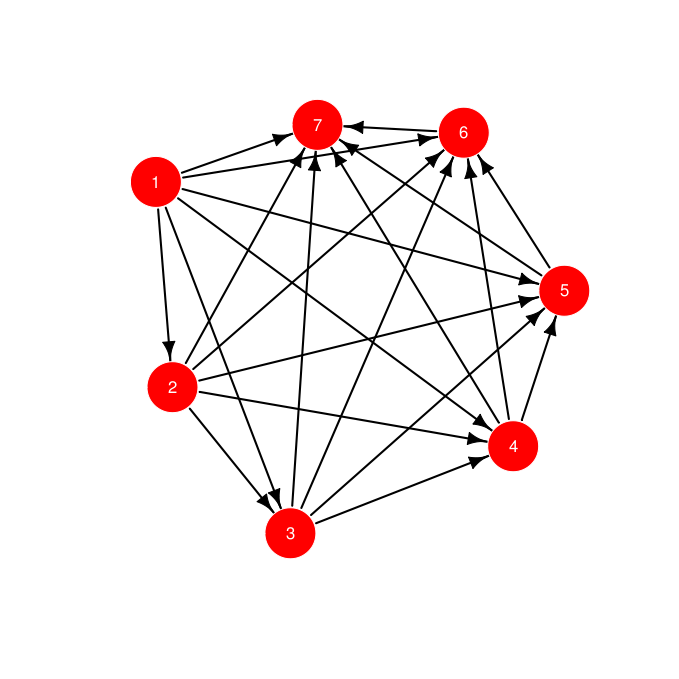}\label{fig: full_graph}}
\subfigure[Sparse graph]{\includegraphics[width=2.5in,height=2.5in]{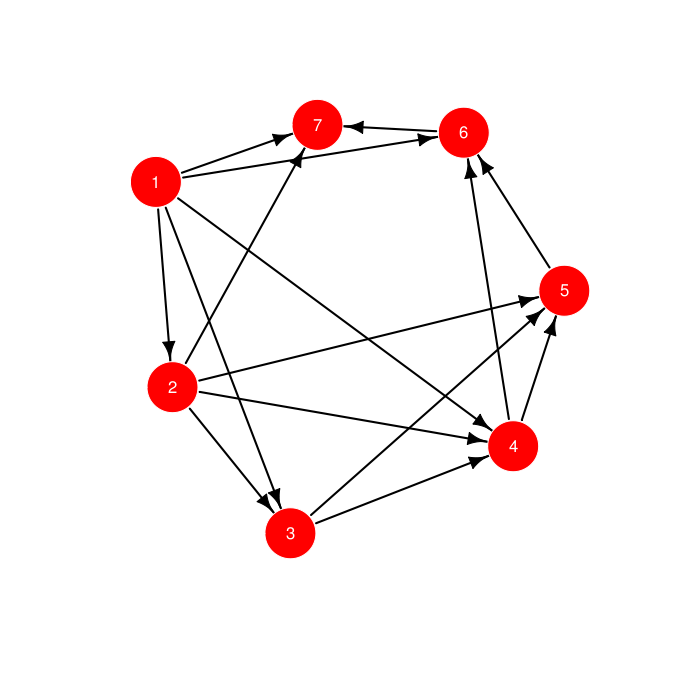}\label{fig: sparse_graph}}
\end{center}
\caption{Sparsity using directed acyclic graphs} \label{fig: graph}
\end{figure}

The underlying idea is, in fact, ubiquitous in graphical models or Bayesian networks \citep[see, e.g.,][]{lauritzen96,bishop2006,murphy2012}. The joint distribution for a random vector $w$ can be looked upon as a directed acyclic graph (DAG) where each node is a random variable $w_i$. We write the joint distribution as 
\[
 p(w_1)\prod_{i=2}^n p(w_i \given w_1,\ldots, w_{i-1}) = \prod_{i=1}^n p(w_i \given w_{\mbox{Pa}[i]})\;,
\]
where $\mbox{Pa}[1]$ is the empty set and $\mbox{Pa}[i] = \{1,2,\ldots,i-1\}$ for $i=2,3,\ldots,n-1$ is the set of parent nodes with directed edges to $i$. This model is specific to the ordering (sometimes called ``topological ordering'') of the nodes. The DAG corresponding to this factorization is shown in Figure~\ref{fig: full_graph} for $n=7$ nodes. One can refer to this as the full graphical model since $\mbox{Pa}[i]$ comprises all nodes preceding $i$ in the topological order. Shrinking $\mbox{Pa}[i]$ from the set of all nodes preceding $i$ to a smaller subset of parent nodes yields a different, but still valid, joint distribution. In spatial settings, each of the nodes in the DAG have associated spatial coordinates. Thus, the parents for any node $i$ can be chosen to include a certain fixed number of ``nearest neighbors'', say based upon their distance from node $i$. For example, Figure~\ref{fig: sparse_graph} shows the DAG when some of the edges are deleted so as to retain at most $3$ nearest neighbors in the conditional probabilities. The resulting joint density is
\begin{align*}
& p(w_1)\times p(w_2\given w_1)\times p(w_3\given w_1,w_2)\times p(w_4\given w_1,w_2,w_3) \times p(w_5\given \cancel{w_1}, w_2,w_3,w_4) \\
&\qquad \times p(w_6\given w_1, \cancel{w_2},\cancel{w_3}, w_4,w_5) \times p(w_7\given w_1,w_2,\cancel{w_3},\cancel{w_4},\cancel{w_5}, w_6)\; . 
\end{align*}
The above model posits that any node $i$, given its parents, is conditionally independent of any other node that is neither its parent nor its child. 

Applying the above notion to multivariate Gaussian densities evinces the connection between conditional independence in DAGs and sparsity. Consider an $n\times 1$ random vector $w$ distributed as $N(0,K_{\theta})$. Writing $N(w\given 0, K_{\theta})$ as $p(w_1)\prod_{i=2}^{n}p(w_i\given w_1, w_2,\ldots, w_{i-1})$ is equivalent to the following set of linear models,
\begin{align*}
 w_1 &= 0 + \eta_1\; \quad \mbox{ and }\; \quad w_i = a_{i1}w_1 + a_{i2}w_2 +  \cdots + a_{i,i-1}w_{i-1} + \eta_i\; \mbox{ for } i=2,\ldots,n\; ,
\end{align*}
or, more compactly, simply $w = Aw + \eta$, where $A$ is $n\times n$ strictly lower-triangular with elements $a_{ij} = 0$ whenever $j \geq i$ and $\eta \sim N(0, D)$ and $D$ is diagonal with diagonal entries $d_{11} = \mbox{var}\{w_1\}$ and $d_{ii} = \mbox{var}\{w_i\given w_j : j < i\}$ for $i=2,\ldots,n$. 

From the structure of $A$ it is evident that $I-A$ is nonsingular and $K_{\theta} = (I-A)^{-1}D(I-A)^{-\top}$. The possibly nonzero elements of $A$ and $D$ are completely determined by the matrix $K_{\theta}$. Let $\texttt{a[i,j]}$, $\texttt{d[i,j]}$ and $\texttt{K[i,j]}$ denote the $(i,j)$-th entries of $A$, $D$ and $K_{\theta}$, respectively. Note that $\texttt{d[1,1] = K[1,1]}$ and the first row of $A$ is $0$. A pseudo-code to compute the remaining elements of $A$ and $D$ is:
\begin{align}\label{eq: pseudocode_full_gaussian}
\begin{array}{ll}
&\texttt{for(i in 1:(n-1))} \texttt{ \{ } \\ 
&\qquad\texttt{a[i+1,1:i] = solve(K[1:i,1:i], K[1:i,i+1])} \\
&\qquad\texttt{d[i+1,i+1] = K[i+1,i+1] - dot(K[i+1,1:i],a[i+1,1:i])}  \\
& \texttt{\}.}
\end{array}
\end{align}
Here  $\texttt{a[i+1,1:i]}$ is the $1\times \texttt{i}$ row vector comprising the possibly nonzero elements of the $\texttt{i+1}$-th row of $A$, $\texttt{K[1:i,1:i]}$ is the $\texttt{i}\times \texttt{i}$ leading principal submatrix of $K_{\theta}$, $\texttt{K[1:i, i]}$ is the $\texttt{i}\times 1$ row vector formed by the first $\texttt{i}$ elements in the $\texttt{i}$-th column of $K_{\theta}$,  $\texttt{K[i, 1:i]}$ is the $1\times \texttt{i}$ row vector formed by the first $\texttt{i}$ elements in the $\texttt{i}$-th row of $K_{\theta}$, $\texttt{solve(B,b)}$ computes the solution for the linear system $\texttt{Bx = b}$, and $\texttt{dot(u,v)}$ provides the inner product between vectors $\texttt{u}$ and $\texttt{v}$. The determinant of $K_{\theta}$ is obtained with almost no additional cost: it is simply $\prod_{\texttt{i=1}}^{\texttt{n}}\texttt{d[i,i]}$.

The above pseudocode provides a way to obtain the Cholesky decomposition of $K_{\theta}$. If $K_{\theta} = LDL^{\top}$ is the Cholesky decomposition, then $L = (I-A)^{-1}$. There is, however, no apparent gain to be had from the preceding computations since one will need to solve increasingly larger linear systems as the loop runs into higher values of $\texttt{i}$. Nevertheless, it immediately shows how to exploit sparsity if we set some of the elements in the lower triangular part of $A$ to be zero. For example, suppose we set at most $\texttt{m}$ elements in each row of $A$ to be nonzero. Let $\texttt{N[i]}$ be the set of indices $\texttt{j} < \texttt{i}$ such that $\texttt{a[i,j]} \neq 0$. We can compute the nonzero elements of $A$ and the diagonal elements of $D$ much more efficiently as:
\begin{align}\label{eq: pseudocode_sparse_gaussian}
\begin{array}{ll}
&\texttt{for(i in 1:(n-1)} \texttt{ \{ } \\
&\qquad \texttt{Pa = N[i+1] \# neighbors of i+1}\\
&\qquad\texttt{a[i+1,Pa] = solve(K[Pa,Pa], K[(i+1),Pa])} \\
&\qquad\texttt{d[i+1,i+1] = K[i+1,i+1] - dot(K[(i+1),Pa], a[i+1,Pa])}\\
& \texttt{\}.}
\end{array}
\end{align}
In (\ref{eq: pseudocode_sparse_gaussian}) we solve $\texttt{n-1}$ linear systems of size at most $\texttt{m}\times \texttt{m}$. This can be performed in $\sim \texttt{nm}^3$ flops, whereas the earlier pseudocode in (\ref{eq: pseudocode_full_gaussian}) for the dense model required $\sim \texttt{n}^3$ flops. These computations can be performed in parallel as each iteration of the loop is independent of the others. 

\begin{figure}[t]
\begin{center}
\subfigure[$I-A$]{\includegraphics[width=3.5cm,trim={2.4cm 2.4cm 0cm 0cm},clip]{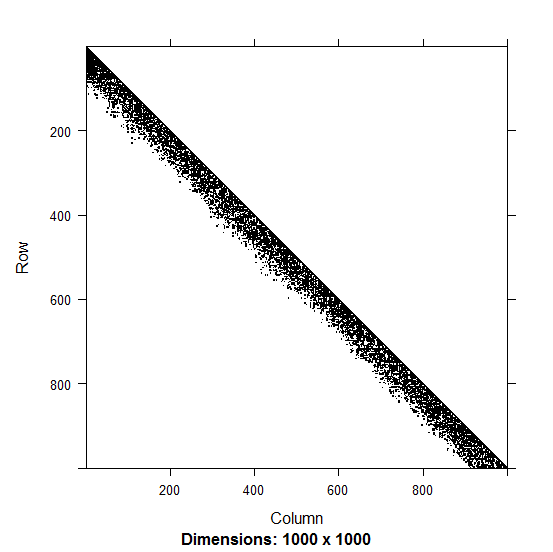}\label{fig: I-A}}
\subfigure[$D^{-1}$]{\includegraphics[width=3.5cm,trim={2.4cm 2.4cm 0cm 0cm},clip]{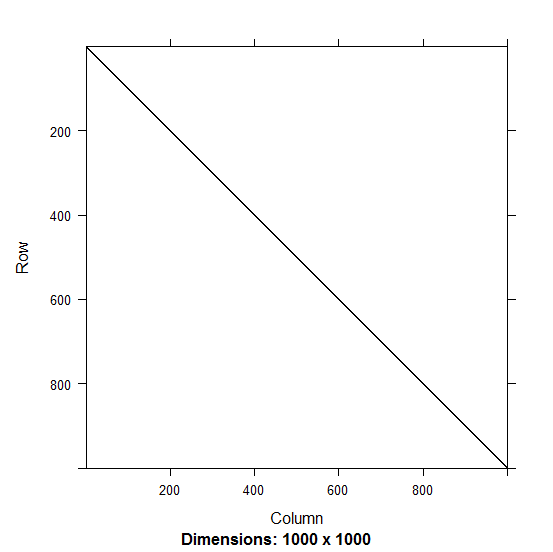}\label{fig: D.inv}}
\subfigure[$\tildeK_{\theta}^{-1}$]{\includegraphics[width=3.5cm,trim={2cm 2cm 0cm 0cm},clip]{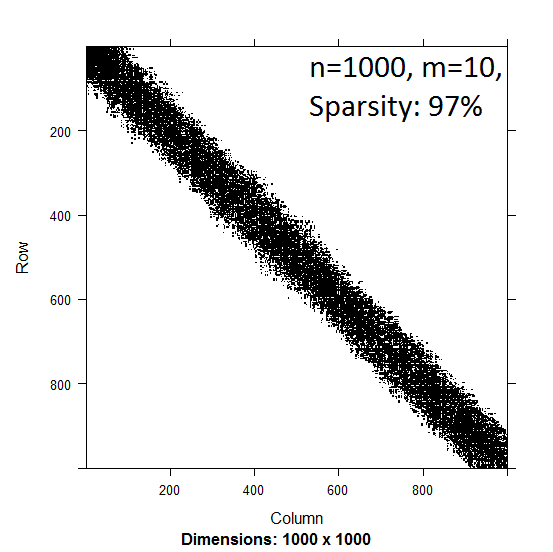}\label{fig: K.theta.inv}}
\end{center}
\caption{Structure of the factors making up the sparse $\tildeK^{-1}_{\theta}$ matrix.} \label{fig: NNGP_Chol}
\end{figure}

The above discussion provides a very useful strategy for introducing sparsity in a precision matrix. Let $K_{\theta}$ and $K^{-1}_{\theta}$ both be dense $n\times n$ positive definite matrices. Suppose we use the pseudocode in (\ref{eq: pseudocode_sparse_gaussian}) with $\texttt{K} = K_{\theta}$ to construct a sparse strictly lower-triangular matrix $A$ with no more than $m$ non-zero entries in each row, where $m$ is considerably smaller than $n$, and the diagonal matrix $D$.  The resulting matrix $\tilde{K}_{\theta} = (I-A)^{-1}D(I-A)^{-\top}$ is a covariance matrix whose inverse $\tilde{K}^{-1}_{\theta} = (I-A^{\T})D^{-1}(I-A)$ is sparse. Figure~\ref{fig: NNGP_Chol} presents a visual representation of the sparsity. While $\tilde{K}_{\theta}$ need not be sparse, the density $N(w\given 0, \tilde{K}_{\theta})$ is cheap to compute since $\tilde{K}^{-1}_{\theta}$ is sparse and $\det(\tilde{K}_{\theta}) = \det(D) = \prod_{i=1}^n\texttt{d[i,i]}$ is calculated from (\ref{eq: pseudocode_sparse_gaussian}). Therefore, one way to achieve massive scalability for models such as (\ref{eq: Bayesian_Spatial_Gaussian_Generic}) is to assume that $w$ has prior $N(w\given 0, \tildeK_{\theta})$ instead of $N(w\given 0, K_{\theta})$.  

\subsection{From distributions to processes}\label{sec: nngp}

If we are interested in estimating the spatial or spatiotemporal process parameters from a finite collection of random variables, then we can use the approach in Section~\ref{sec: sparse_ggm} with $w_i := w(\ell_i)$. In spatial settings, matters are especially convenient as we can delete the edges in the DAG based upon the distances among $\ell_i$'s. In fact, one can decide to retain at most $m$ of the nearest neighbors for each location and delete all remaining edges. This implies that the $(i,j)$-th element of $A$ in Section~\ref{sec: sparse_ggm} will be nonzero only if $\ell_j$ is one of the $m$ nearest neighbors of $\ell_i$. In fact, this idea has been effectively used to construct composite likelihoods for Gaussian process models by \cite{ve88} and \cite{stein04}, while \cite{stroud17} exploits this idea to propose preconditioned conjugate gradient algorithms for Bayesian and maximum likelihood estimates on large incomplete lattices.  

Localized Gaussian process regression based on few nearest neighbors has also been used to obtain fast kriging estimates. \cite{emery09} provides fast updates for kriging equations after adding a new location to the input set. Iterative application of their algorithm yields a localized kriging estimate based on a small set of locations (including few nearest neighbors). The local estimate often provides an excellent approximation to the global kriging estimate which uses data observed at all the locations to predict at a new location. However, this assumes that the parameters associated with the mean and covariance of the GP are known or already estimated. Local Approximation GP, or LAGP \citep{gram14,gramhaal2016,gram2016}, extends this further to estimate the parameters at each new location,  essentially  providing a non-stationary local approximation to a Gaussian Process at every predictive location and can be used to interpolate or smooth the observed data.

If, however, posterior predictive inference is sought at arbitrary spatiotemporal resolutions, i.e., for the entire process $\{w(\ell) : \ell \in \calL\}$, then the ideas in Section~\ref{sec: sparse_ggm} need to be extended to process-based models. Recently, \cite{datta16} proposed a Nearest Neighbor Gaussian Process (NNGP) for modeling large spatial data. NNGP is a well defined Gaussian Process over a domain $\calL$ and yields finite dimensional Gaussian densities with sparse precision matrices. This has been also extended to a dynamic NNGP with dynamic neighbor selection for massive spatiotemporal data \citep[][]{datta16b}. The NNGP delivers massive scalability both in terms of parameter estimation and kriging. Unlike low rank processes, it does not oversmooth and accurately emulates the inference from full rank GPs. 

We will construct the NNGP in two steps. First, we specify a multivariate Gaussian distribution over a fixed finite set $r$ points in $\calL$, say $\calR = \{\ell_1^*, \ell_2^*,\ldots,\ell_r^*\}$, which we call the \emph{reference set}. The reference set can be very large. It can be a fine grid of points over $\calL$ or one can simply take $r=n$ and let $\calR$ be the set of observed points in $\calL$. We require that the inverse of the covariance matrix be sparse and computationally efficient. Therefore, we specify that $w_{\calR} \sim N(0, \tilde{K}_{\theta})$, where $w_{\calR}$ is the $r\times 1$ vector with elements $w(\ell_i^*)$ and $\tilde{K}_{\theta}$ is a covariance matrix such that $\tildeK^{-1}_{\theta}$ is sparse. The matrix $\tildeK_{\theta}$ is constructed from a dense covariance matrix $K_{\theta}$ as described in Section~\ref{sec: sparse_ggm}. This provides a highly effective approximation \citep[][]{ve88,stein04} as below:      
\begin{align}\label{eq: vecchia_approx}
 N(w_{\calR}\given 0, K_{\theta}) &= \prod_{i=1}^r p(w(\ell^*_{i}) \given w_{H(\ell^*_{i})}) \approx \prod_{i=1}^r p(w(\ell^*_{i}) \given w_{N(\ell^*_{i})}) = N(w_\calR \given 0, \tildeK_{\theta})\; ,
\end{align}
where \emph{history sets} $H(\ell_{i}^*)$ so that $H(\ell^*_1)$ is the empty set and $H(\ell^*_i) = \{\ell^*_1, \ell^*_2,\ldots, \ell^*_{i-1}\}$ for $i=2,3,\ldots,r$ and we have much smaller \emph{neighbor sets} $N(\ell_{i}^*)\subseteq H(\ell_i^*)$ for each $\ell_{i}^*$ in $\calR$. We have legitimate probability models for any choice of $N(\ell_i^*)$'s as long as $N(\ell_{i}^*)\subseteq H(\ell_i^*)$. One easy specification is to define $N(\ell^*_i)$ as the set of $m$ nearest neighbors of $\ell_i^*$ among the points in $\calR$. Therefore,
\begin{align*}
 N(\ell_i) = \left\{\begin{array}{l}
 		       \mbox{ empty set for } i=1 \\      
                H(\ell_i^*) =\{\ell^*_1,\ell^*_2,\ldots,\ell^*_{i-1}\}\; \mbox{ for }\;  i=2,3,\ldots, m \\
		  	   m \mbox{ nearest neighbors of $\ell^*_i$ among } H(\ell_i^*) \; \mbox{ for } i = m+1,\ldots,n
              \end{array}
 \right.\; .
\end{align*}
If $m (<< r)$ denotes the limiting size of the neighbor sets $N(\ell)$, then $\tildeK^{-1}_{\theta}$ has at most $O(rm^2)$ non-zero elements. Hence, the approximation in (\ref{eq: vecchia_approx}) produces a sparsity-inducing proper prior distribution for random effects over $\calR$ that closely approximates the realizations from a $GP(0, K_{\theta})$. 

To construct the NNGP we extend the above model to arbitrary locations. We define neighbor sets $N(\ell)$ for any $\ell\in \calL$ as the set of $m$ nearest neighbors of $\ell$ in $\calR$. Thus, $N(\ell)\subseteq \calR$ and the process can be derived from $p(w_{\calR}, w(\ell)\given \theta) =  N(w_{\calR} \given 0, \tildeK_{\theta})\times p\left(w(\ell) \given w_{N(\ell)},\theta\right)$ or, equivalently, by writing
\begin{equation}\label{eq:sgp}
w(\ell) = \sum_{i=1}^r a_{i}(\ell) w(\ell_i^*)  + \eta(\ell) \mbox{ for any } \ell \notin \calR\; ,
\end{equation}
where $a_i(\ell) = 0$ whenever $\ell_i^* \notin N(\ell)$, $\eta(\ell) \stackrel{ind}{\sim} N(0, \delta^2(\ell))$ is a process independent of $w(\ell)$, $\mbox{Cov}\{\eta(\ell), \eta(\ell')\}=0$ for any two distinct points in $\calL$, and $$\delta^2(\ell) = K_{\theta}(\ell,\ell) - K_{\theta}(\ell,N(\ell))K_{\theta}^{-1}(N(\ell),N(\ell))K_{\theta}(N(\ell), \ell)\;.$$ Taking conditional expectations in (\ref{eq:sgp}) yields 
$
\mbox{E}[w(\ell)\given w_{N(\ell)}] = \sum_{i: \ell_i\in N(\ell)}a_i(\ell)w(\ell_i^*)\; ,
$
which implies that for each $\ell$ the nonzero $a_i(\ell)$'s are obtained by solving an $m\times m$ linear system. The above construction ensures that $w(\ell)$ is a legitimate Gaussian process whose realizations over any finite collection of arbitrary points in $\calL$ will have a multivariate normal distribution with a sparse precision matrix. More formal developments and technical details in the spatial and spatiotemporal settings can be found in \cite{datta16} and \cite{datta16b}, respectively. 

One point worth considering is the definition of ``neighbors.'' There is some flexibility here. In the spatial setting, the correlation functions usually decay with increasing inter-site distance, so the set of nearest neighbors based on the inter-site distances represents locations exhibiting highest correlation with the given locations. For example, on the plane one could simply use the Euclidean metric to construct neighbor sets, although \cite{stein04} recommends including a few points that are farther apart. The neighbor sets can be fixed before the model fitting exercise. 

In spatiotemporal settings, matters are more complicated. Spatiotemporal covariances between two points typically depend on the spatial as well as the temporal lag between the points. Non-separable isotropic spatiotemporal covariance functions can be written as $K_{\theta}((s_1,t_1),(s_2,t_2)) = K_{\theta}(h,u)$ where $h=\|s_1-s_2 \| $ and $u=|t_1-t_2|$. This often precludes defining any universal distance function $d: (\calS \times \calT)^2 \rightarrow \Re^+$ such that $K_{\theta}((s_1,t_1),(s_2, t_2))$ will be monotonic with respect to $d((s_1,t_1),(s_2,t_2))$ for all choices of $\theta$. This makes it difficult to define universal nearest neighbors in spatiotemporal domains. To obviate this hurdle, \cite{datta16b} define ``nearest neighbors'' in a spatiotemporal domain using the spatiotemporal covariance function itself as a proxy for distance. This can work for arbitrary domains. For any three points $\ell_1$, $\ell_2$ and $\ell_3$, we say that $\ell_1$ is nearer to $\ell_2$ than to $\ell_3$ if $K_{\theta}(\ell_1,\ell_2) > K_{\theta}(\ell_1,\ell_3)$. Subsequently, this definition of ``distance'' is used to find $m$ nearest neighbors for any location. 
Prediction at any arbitrary location $\ell\notin\calR$ is performed by sampling from the posterior predictive distribution.
However, for every point $\ell_i$, its neighbor set $N_\theta(\ell)$ will now depend on $\theta$ and can change from iteration to iteration in the estimation algorithm. If $\theta$ were known, one could have simply evaluated the pairwise correlations between any point $\ell_i^*$ in $\calR$ and all points in its history set $H(\ell_i^*)$ to obtain $N_\theta(\ell_i^*)$ --- the set of $m$ true nearest neighbors. In practice, however, $\theta$ is unknown and for every new value of $\theta$ in an iterative algorithm, we need to search for the neighbor sets within the history sets. Since the history sets are quite large, searching the entire space for nearest neighbors in each iteration will be computationally unfeasible. \cite{datta16b} offer some smart strategies for selecting spatiotemporal neighbors. They propose restricting the search for the neighbor sets to carefully constructed small subsets of the history sets. These small {\em eligible sets} $E(\ell_i^*)$ are constructed in such a manner that, despite being much smaller than the history sets, they are guaranteed to contain the true nearest neighbor sets. This strategy works when we choose $m$ to be a perfect square and the original nonseparable covariance function $K_{\theta}(h,u)$ satisfies {\em natural monotonicity}, i.e. $K_{\theta}(h,u)$ is decreasing in $h$ for fixed $u$ and decreasing in $u$ for fixed $h$. All Mat\`ern-based space-time separable covariances and many non-separable classes of covariance functions possess this property \citep{stein2013,omidi15}.

\subsection{Hierarchical NNGP models}\label{sec: hierarchical_nngp_models}

We briefly turn to model fitting and estimation. For the approximation in (\ref{eq: vecchia_approx}) to be effective, the size of the reference set, $r$, needs to be large enough to represent the spatial domain. However, this does not impede computations involving NNGP models because the storage and number of floating point operations are always linear in $r$. The reference set $\calR$ can, in principle, be any finite set of locations in the study domain. A particularly convenient choice, in practice, is to simply take $\calR$ to be the set of observed locations in the dataset. \cite{datta16} demonstrate through extensive simulation experiments and a real application that this simple choice seems to be very effective. 

Since the NNGP is a proper Gaussian process, we can use it as a prior for the spatial random effects in any hierarchical model. We write $w(\ell) \sim NNGP(0, \tildeK_{\theta}(\cdot,\cdot))$, where $\tildeK_{\theta}(\ell,\ell')$ is the covariance function for the NNGP \citep[see][for a closed form expression]{datta16}. For example, with $r=n$ and $\calR$ the set of observed locations, one can build a scalable Bayesian hierarchical model exactly as with a usual spatial process, but assigning an NNGP to the spatial random effects. Here is a simple NNGP-based spatial model with a first stage exponential family model:
\begin{align}\label{eq: nngp_latent}
\begin{array}{cll}
  Y(\ell) \given g(\cdot), \beta, w(\ell) &\stackrel{ind}{\sim} P_{\tau}\quad \mbox{ exponential family}\;, \\
   g(\mbox{E}[Y(\ell)]) &= x^{\T}(\ell)\beta + w(\ell)\;,\quad  w(\ell) \sim NNGP(0, \tildeK_{\theta}(\cdot,\cdot))\;, \\ 
   \{\theta, \beta, \tau\} &\sim p(\theta, \beta, \tau)\; ,
\end{array}
\end{align}
where $P_{\tau}$ is an exponential family distribution with link function $g(\cdot)$.
Posterior sampling from (\ref{eq: nngp_latent}) is customarily performed using Gibbs sampling with Metropolis steps. Computational benefits emerge from the fact that the full conditional distribution $p(w(\ell_i)\given w_{\calR}, \theta, \beta, \tau) = p(w(\ell_i)\given w_{N(\ell_i)}, \theta, \beta, \tau)$ and since $w_{N(\ell_i)}$ is an $m\times 1$ subset of $w_{\calR}$. Prediction at any arbitrary location $\ell\notin\calR$ is performed by sampling from the posterior predictive distribution. For each draw of  $\{w_{\calR}, \beta, \theta, \tau\}$ from $p(w_{\calR}, \beta, \tau, \theta \given y)$, we draw a $w(\ell)$ from $N(a^{\T}(\ell)w_{N(\ell)}, \delta^2(\ell))$ and $y(\ell)$ from $p(y(\ell) \given \beta, w(\ell),\tau)$, where $y$ is the vector of observed outcomes and $a(\ell)$ is a vector of the nonzero $a_j(\ell)$'s in (\ref{eq:sgp}).

Another, even simpler, example could be modeling a continuous outcome itself as an NNGP. Let the desired full GP specification be $Y(\ell)\sim GP(x^{\top}(\ell)\beta, K_{\theta}(\cdot, \cdot))$. We derive the NNGP from this $K_{\theta}$ and obtain
\begin{align}\label{eq: nngp_outcome}
   Y(\ell) &\sim NNGP(\mu(\ell), \tildeK_{\theta}(\cdot,\cdot))\;; \quad \mu(\ell) = x^{\T}(\ell)\beta\;;\quad  \{\theta,\beta\} \sim p(\theta, \beta)\; .
\end{align}
The above model is extremely fast. The likelihood is of the form $y\sim N(X\beta, \tildeK_{\theta})$, where $\tilde{K}^{-1}_{\theta} = (I-A^{\T})D^{-1}(I-A)$ is sparse and $A$ and $D$ are obtained from (\ref{eq: pseudocode_sparse_gaussian}) efficiently in parallel. The parameter space of interest is $\{\theta,\beta\}$, which is much smaller than for (\ref{eq: nngp_latent}) where the latent spatial process also was unknown. While (\ref{eq: nngp_outcome}) does not separate the residuals into a spatial process and a measurement error process, one can still include measurement error variance, or the nugget, in (\ref{eq: nngp_outcome}). Here, one would absorb the nugget into $\theta$. For example, we could write the likelihood in (\ref{eq: BasicModel}) as $N(y\given X\beta, K_{\theta})$, where $K_{\theta} = \sigma^2 R_{\phi} + \taus I_n$, $R_{\phi}$ is a spatial correlation matrix and $\theta = \{\sigma^2,\phi,\tau^2\}$. These will also feature in the derived NNGP covariance matrix $\tildeK_{\theta}$. We can predict the outcome at an arbitrary point $\ell$ by sampling from the posterior predictive distribution as follows: for each draw of $\{\beta, \theta\}$ from $p(\beta, \theta \given y)$, we draw a $y(\ell)$ from $N(y(\ell) \given x^{\T}(\ell)\beta, \delta^2(\ell))$. Note, however, that there is no latent smooth process $w(\ell)$ in (\ref{eq: nngp_outcome}) and inference on the latent spatial process is precluded.

\begin{figure}[t]
 \begin{center}
  \includegraphics[width=3.5in]{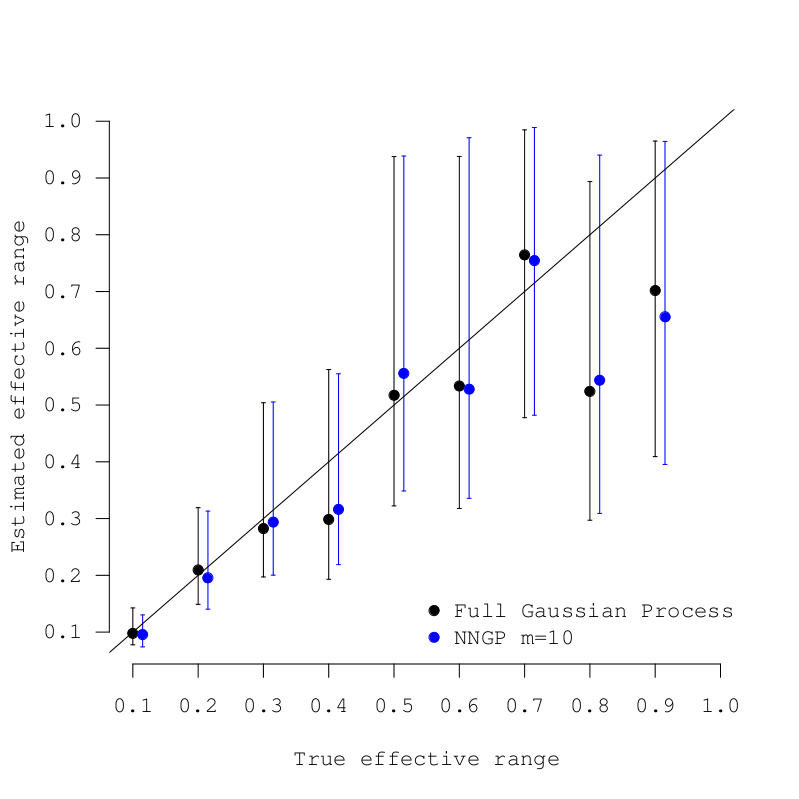}
 \end{center}
 \caption{95\% credible intervals for the effective spatial range from an NNGP model with $m=10$ and a full GP model fitted to 10 different simulated datasets with true effective range fixed at values between 0.1 and 1.0 in increments of 0.1.}\label{fig: effective_range}
\end{figure}

Likelihood computations in NNGP models usually involve $O(nm^3)$ flops. One does not need to store $n\times n$ matrices, only $m\times m$ matrices which leads to storage $\sim nm^2$. Substantial computational savings accrue because $m$ is usually very small.
\cite{datta16} demonstrate that fitting NNGP models to the simulated data in Figure~\ref{fig:uni-w} with number of neighbors as less as $m=10$ produce posterior estimates of the spatial surface indistinguishable from Figures~\ref{uni-w-obs}~and~\ref{uni-w-gs}. In fact, simulation experiments in \cite{datta16} and \cite{datta16b} also affirm that $m$ can usually be taken to be very small compared to $r$; there seems to be no inferential advantage to taking $m$ to exceed 15, even for datasets with over $10^5$ spatial locations. For example, Figure~\ref{fig: effective_range} shows the $95\%$ posterior credible intervals for a series of 10 simulation experiments where the true effective range was fixed at values from 0.1 to 1.0 in increments of 0.1. Each dataset comprised $2500$ points. Even with $m=10$ neighbors, the credible intervals for the effective spatial range from the NNGP model were very consistent with those from the full GP model. \cite{datta16} present simulations using the Mat\'ern and other covariance functions revealing very similar behavior.

Another important point to note is that $\tildeK_{\theta}$ is not invariant to the order in which we define $H(\ell_1) \subseteq H(\ell_2) \subseteq \cdots \subseteq H(\ell_r)$ (i.e., the topological order). \cite{ve88} and \cite{stein04} both assert that the approximation in (\ref{eq: vecchia_approx}) is not sensitive to this ordering. This is corroborated by simulation experiments by \cite{datta16}, but a recent manuscript by \cite{guinness16} has indicated sensitivity to the ordering in terms of model deviance. We conducted some preliminary investigations to investigate the effect of the topological order. In one simple experiment we generated data from the ``true'' model in (\ref{eq: BasicModel}) for $6400$ spatial locations arranged over an $80\times 80$ grid. The parameter $\beta$ in (\ref{eq: BasicModel}) was set to $0$, the covariance function was specified as $K_{\theta}(\ell_i,\ell_j) = \sigma^2 \exp(-\phi \|\ell_i-\ell_j\|)$, and $\epsilon(\ell_i) \stackrel{iid}{\sim} N(0,\tau^2)$ with the true values of $\sigma^2$, $\phi$ and $\taus$ given in the second column of Table~\ref{tab:order}. Four different NNGP models corresponding to (\ref{eq: nngp_outcome}) with $\tildeK_{\theta}$ derived from $K_{\theta} = \sigma^2R_{\phi} + \taus I$ and $R_{\phi}$ having elements $\exp(-\phi \|\ell_i-\ell_j\|)$, were fitted to the simulated data. Each of these models were constructed with $m=10$ nearest neighbors, but with different ordering of the points $\ell = (x,y)$. These were performed according to the sum of the coordinates $x+y$, a maximum-minimum distance (MMD) proposed by \cite{guinness16}, the $x$ coordinate, and the $y$ coordinate. Table~\ref{tab:order} presents a comparison of these NNGP models. Irrespective of the ordering of the points, the inference with respect to parameter estimates and predictive performance is extremely robust and effectively indistinguishable from each other. However, the posterior mean of the Kullback-Leibler divergence of these models from the true generating model revealed that the metric proposed by \cite{guinness16} is indeed less than the other three. Further explorations are currently being conducted to see how this behavior changes for more complex nonstationary models and in more general settings.   

\begin{table} 
	\begin{center} 
	\ra{1.3} 
	\resizebox{\textwidth}{!}{\begin{tabular}{@{}lllllll@{}}
				\toprule
				\multicolumn{6}{c} {NNGP from different topological orders}\\
				\toprule
				& True & Sorted coord(x+y)& MMD & Sorted x & Sorted y\\
				$\sigma$ &1 &0.79 (0.69, 1.04 ) & 0.80 (0.69, 1.02)  &   0.80 (0.70, 1.05) & 0.83 (0.69, 1.08) \\
				$\tau$ &0.45 & 0.45 (0.44, 0.46) & 0.45 (0.44, 0.47)& 0.45 (0.44, 0.46 )& 0.45 (0.44, 0.47) \\
				$\phi$ & 5 &  8.11 (4.42, 11.10) &  7.63 (4.58, 10.97) & 8.01 (4.26,  11.18)  & 7.12 (4.06, 11.03) \\
				\hline
				KL-D & --& 24.04022&13.88847 & 22.30667 & 21.59174\\
				RMSPE&-- &0.5278996 & 0.5278198&0.527912 &0.527807 \\
				\bottomrule
			\end{tabular}}
			\caption{Posterior parameter estimates, the Kullback-Leibler divergence (KL-D) and root mean square predictive errors (RMSPE) are presented for four NNGP models constructed from different topological orderings. The four orderings from left to right are ``sorted on the sum of vertical and horizontal coordinate'', maximum-minimum distance \citep[][]{guinness16}, sorted on horizontal coordinate and sorted on vertical coordinate.} \label{tab:order}
	\end{center}
\end{table}

\section{Discussion and future directions}\label{sec:future_directions}
The article has attempted to provide some insight into constructing highly scalable Bayesian hierarchical models for very large spatiotemporal datasets using low-rank and sparsity-inducing processes. Such models are increasingly being employed to answer complex scientific questions and analyze massive spatiotemporal datasets in the natural and environmental sciences. Any standard Bayesian estimation algorithm, such as Markov chain and Hamiltonian Monte Carlo \citep[see, e.g.,][]{robecase04,brooksgelmanjonesmeng2011,gelman2013,neal2011,hoffmangelman2014}, Integrated Nested Laplace Approximations \citep[][]{ruemartinochopin2009}, and Variational Bayes \citep[see, e.g.,][]{bishop2006} can be used for fitting these models. The models ensure that the algorithmic complexity has $\sim n$ floating point operations (flops), where $n$ the number of spatial locations (per iteration). Storage requirements are also linear in $n$. Methods such as the multiresolution predictive process \citep[][]{katzfussmultires} and the NNGP \citep[][]{datta16} can scale up to datasets in the order of $\sim 10^6$ spatial and/or temporal points without sacrificing richness in the model. 

While the NNGP certainly seem to have an edge in scalability over the more conventional low-rank or fixed rank models, it is premature to say whether its inferential performance will always excel over low rank of fixed rank models. For example, analyzing complex nonstationary random fields may pose challenges regarding construction of neighbor sets as simple distance-based definition of neighbors may prove to be inadequate. Multiresolution basis functions may be more adept at capturing nonstationary, but may struggle with massive datasets. Dynamic neighbor selection for nonstationary fields, where neighbors will be chosen based upon the covariance kernel itself, analogous to \cite{datta16b} for space-time covariance functions, may be an option worth exploring. Multiresolution NNGPs, where the residual from the NNGP approximation is modeled hierarchically \citep[analogous to][for the predictive process]{katzfussmultires} may also be promising in terms of full Bayesian inference at massive scales.    

There remain other challenges in high-dimensional geostatistics. Here, we have considered geostatistical settings where we have very large numbers of locations and/or time-points, but restricted our discussion to univariate outcomes. In practice, we often observe a $q \times 1$ variate response $y(\ell)$ along with a set of explanatory variables $X(\ell)$ and $q\times 1$ variate GP, $w(\ell)$, is used to capture the spatial patterns beyond the observed covariates. We seek to capture associations among the variables as well as the strength of spatiotemporal association for each outcome. One specific geostatistical problem in ecology that currently lacks a satisfying solution is a joint species distribution model, where we seek to model a large collection of species (say, order $10^3$) over a large collection of spatial sites (again, say, order $10^3$).  

The linear model of coregionalization (LMC) proposed by \citet{Matheron1982} is among the most general models for multivariate spatial data analysis. Here, the spatial behavior of the outcomes is assumed to arise from a linear combination of the independent latent processes operating at different spatial scales \citep{Chiles1999}. The idea resembles latent factor analysis (FA) models for multivariate data analysis \citep[e.g.,][]{Anderson2003} except that in the LMC the number of latent processes is usually taken to be the same as the number of outcomes. Then, an $q \times q$ covariance matrix has to be estimated for each spatial scale \citep[see, e.g.,][]{Lark2003, Castrignano2005, Zhang2007}, where $q$ is the number of outcomes. When $q$ is large (e.g., $q \geq 5$ and 300 spatial locations), obtaining such estimates is expensive. \citet{Schmidt2003} and \citet{Gelfand2004} associate only a $q \times q$ triangular matrix with the latent processes. However, high dimensional outcomes are still computationally prohibitive for these models. 

Spatial factor models \citep[see, e.g.,][]{Lopes2004,Lopes2008,Wang2003} have been used to handle high dimensional outcomes but with modest number of spatial locations. Dimension reduction is needed in two aspects: (i) the length of the vector of outcomes, and (ii) the very large number of spatial locations. Latent variable (factor) models are usually used to address the former, while low-rank spatial processes offer a rich and flexible modeling option for dealing with a large number of locations. \cite{renbanerjee2013} have exploited these two ideas to propose a class of hierarchical low-rank spatial factor models and also explored stochastic selection of the latent factors without resorting to complex computational strategies (such as reversible jump algorithms) by utilizing certain identifiability characterizations for the spatial factor model. Their model was designed to capture associations among the variables as well as the strength of spatial association for each variable. In addition, they reckoned with the common setting where not all the variables have been observed over all locations, which leads to \emph{spatial misalignment}.  The fully Bayesian approach effectively deals with spatial misalignment, but is likely to suffer from the limited ability of low-rank models to scale to a very large number of locations. Promising ideas include using the multiresolution predictive process or the NNGP as a prior on the spatial factors.  

Computational developments with regard to Markov chain Monte Carlo (MCMC) algorithms \citep[see, e.g.,][]{robecase04} have contributed enormously to the dissemination of Bayesian hierarchical models in a wide array of disciplines. Spatial modeling is no exception. However, the challenges for automated implementation of geostatistical model fitting and inference are substantial. First, expensive matrix computations are required that can become prohibitive with large datasets. Second, routines to fit unmarginalized models are less suited for direct updating using a Gibbs sampler and result in slower convergence of the chains. Third, investigators often encounter multivariate spatial datasets with several spatially dependent outcomes, whose analysis requires multivariate spatial models that involve demanding matrix computations. These issues have, however, started to wane with the delivery of relatively simpler software packages in the \texttt{R} statistical computing environment via the Comprehensive R Archive Network (CRAN) (\url{http://cran.r-project.org}). Several packages that automate Bayesian methods for point-referenced data and diagnose convergence of MCMC algorithms are easily available from CRAN. Packages that fit Bayesian models include \texttt{geoR}, \texttt{geoRglm}, \texttt{spTimer}, \texttt{spBayes}, \texttt{spate}, and \texttt{ramps}. 

In terms of the hierarchical geostatistical models presented in this article, \texttt{spBayes} offers users a suite of Bayesian hierarchical models for Gaussian and non-Gaussian univariate and multivariate spatial data as well as dynamic Bayesian spatio-temporal models. It focuses upon performance issues for full Bayesian inference, sampler convergence rate and efficiency using a collapsed Gibbs sampler, decreasing sampler run-time by avoiding expensive matrix computations, and increased scalability to large datasets by implementing predictive process models. Beyond these general computational improvements for existing models, it analyzes data indexed both in space and time using a class of dynamic spatiotemporal models, and their predictive process counterparts, for settings where space is viewed as continuous and time is taken as discrete. Finally, we have modeling environments such as \texttt{Nimble} \citep[][]{Nimble2017} that gives users enormous flexibility to choose algorithms for fitting their models, and \texttt{Stan} \citep[][]{Stan2017} that estimates Bayesian hierarchical models using Hamiltonian dynamics. The NNGP and the predictive process can be also coded in \texttt{Nimble} and \texttt{Stan} fairly easily.  

\section*{Acknowledgments}
The author wishes to thank Professor Bruno Sans\'{o} and two anonymous reviewers for very constructive and insightful feedback. In addition, the author also wishes to thank Dr. Abhirup Datta, Dr. Andrew O. Finley and Ms. Lu Zhang for useful discussions. The work of the author was supported in part by NSF DMS-1513654 and NSF IIS-1562303. 



\bibliographystyle{bib/ba}
\bibliography{bib/banerjee}


\end{document}